\documentclass[12pt]{article}
\usepackage[utf8]{inputenc}
\usepackage[text={6.5in,9in}]{geometry}
\usepackage{amsthm}

\usepackage{etoolbox}
\BeforeBeginEnvironment{assumption}{\vskip-1ex}
\AfterEndEnvironment{assumption}{\noindent \vskip-1ex}

\usepackage{tikz}
\usetikzlibrary{trees}
\usetikzlibrary{calc} 
\usepackage{authblk,setspace,float}

\title{A Bayesian Multi-Layered Record Linkage Procedure to Analyze Functional Status of Medicare Patients with Traumatic Brain Injury}
\date{}
\author{Mingyang Shan \thanks{Eli Lilly and Company, Indianapolis, IN 46225} , Kali Thomas\thanks{Department of Health Services, Policy, and Practice, Brown University, Providence, RI 02912} , Roee Gutman\thanks{Department of Biostatistics, Brown University, Providence, RI 02912}}

\usepackage{amsthm,amsmath,geometry,textgreek,bbm,amssymb,graphicx,comment,multirow,lscape,hhline,subcaption}
\usepackage[round]{natbib}
\begin{document}
\maketitle
\begin{abstract}
   Understanding the association between injury severity and patients' potential for recovery is crucial to providing better care for patients with traumatic brain injury (TBI). Estimation of this relationship requires clinical information on injury severity, patient demographics, and healthcare utilization, which are often obtained from separate data sources. Because of privacy and confidentiality regulations, these data sources do not include unique identifiers to link records across data sources. Record linkage is a process to identify records that represent the same entity across data sources that contain similar individuals in the absence of unique identifiers. These processes commonly rely on agreement between variables that appear in both data sources to link records. However, when the number of records in each file is large, this task is computationally intensive and may result in false links. Blocking is a data partitioning technique that reduces the number of possible links that should be considered. Healthcare providers can be used as blocks in applications of record linkage with healthcare datasets. However, providers may not be uniquely identified across files. We propose a Bayesian record linkage procedure that simultaneously performs block-level and record-level linkage. This iterative approach incorporates the record-level linkage within block pairs to improve the accuracy of the block-level linkage. Subsequently, the algorithm improves record-level linkage using the accurate partitioning of the linkage space through blocking. We demonstrate that our proposed methodology provides improved performance compared to existing Bayesian record linkage methods that do not incorporate blocking. The proposed procedure is then used to merge registry data from the National Trauma Data Bank with Medicare claims data to estimate the relationship between injury severity and TBI patients' recovery.

\end{abstract}

\section{Introduction}
Traumatic brain injury (TBI) is a leading cause of mortality and disability in the United States, especially among adults over the age of 65 (\cite{Faul2010}, \cite{Gardner2018}). The high incidence rate coupled with evidence of demanding medical and post-acute care poses a growing and potentially costly public health concern (\cite{Thompson2006}, \cite{Dams2013}). Understanding a patient's potential for functional improvement and regaining independence during post-acute care is critical to discharge planning and identifying long term health needs. Heterogeneity in trauma characteristics and pre-injury factors can greatly impact the functional recovery and long term health outcomes of patients (\cite{Yue2013}, \cite{Peters2018}). Previous studies have either examined detailed injury and acute clinical information but focused on short term health outcomes such as mortality (\cite{Clark2007}, \cite{Aitken2010}, \cite{Dinh2013}), or investigated functional outcome assessments without accounting for injury or treatment information (\cite{Buchanan2003}). A major factor contributing to the lack of research that examines how pre-existing medical conditions and functional status impact long term post-acute health outcomes is the lack of a complete dataset that includes detailed injury and clinical data, functional status assessments, patient histories, and long term health outcomes. This data can be obtained by linking records from Medicare enrollment and claims data to records from the National Trauma Data Bank registry. However, data confidentiality restrictions limit the availability of unique identifiers to link records from these two datasets.

Record linkage is the process of identifying entries across two files that represent the same entity. Record linkage methods can be broadly classified into deterministic and probabilistic techniques. Deterministic techniques identify links by examining whether data elements existing in both files agree in value (\cite{Gomatam2002}). These methods yield a high proportion of true links among record pairs that are linked. However, deterministic techniques may miss a large number of true links when the data elements are subject to typographical errors, differences in coding systems, or missingness (\cite{Campbell2008}). Probabilistic record linkage leverages a mixture model proposed by \citet{Fellegi1969} to calculate probabilities that a pair of records from the two datasets represent a true link. This approach commonly employs a greedy algorithm to select the record pairs with the highest probabilities of being true links. Record pairs below a probability cutoff value are either sent for clerical review or classified as non-links. Both deterministic and probabilistic linking methods may utilize optimization procedures to obtain one-to-one linking configurations (\cite{Jaro1989}). These optimization steps enforce a dependency structure among possible links using linking probabilities that are estimated independently. This process may result in suboptimal linking configurations (\cite{Belin1995}). Furthermore, many deterministic and probabilistic methods do not provide a mechanism to quantify the uncertainty in the linking process when analyzing the linked datasets. 

Bayesian record linkage procedures are probabilistic record linkage techniques that were proposed as possible solutions to address these limitations. These Bayesian procedures introduce a missing linkage structure that is sampled together with model parameters. Constraints can be placed on the linkage structure to directly obtain one-to-one linking configurations, eliminating the need for additional post-processing steps. One type of Bayesian record linkage procedure relies on models that describe similarity measures between variables appearing in both files using a mixture of matches and non-matches (\cite{Fortini2001}, \cite{Larsen2005}, \cite{Sadinle2017}). A different set of Bayesian methods relies on measurement error models to describe a set of observed discrete matching variables rather than similarity measures (\cite{Tancredi2011}, \cite{Steorts2015}). Both types of Bayesian methods rely on variables contained in both files, but do not consider variables that appear in only one of the files. \citet{Gutman2013} proposes an approach that utilizes relationships between variables that appear in only one of the files, as well as those appearing in both files to link records. One attractive property of Bayesian record linkage procedures is that they account for errors in the linking process by using samples from the posterior distribution of the linkage structures. These samples can then be used to obtain unbiased point estimates and statistically valid interval estimates (\cite{Gutman2013}). 

Identifying all entities across two datasets can be computationally complex, especially with large numbers of records in each file. Blocking, or indexing, is commonly implemented in record linkage to reduce computation and improve the scalability of linkage algorithms by only considering record pairs that agree on a set of blocking fields (e.g. address, gender, etc.) (\cite{Newcombe1962}, \cite{Fellegi1969}, \cite{Newcombe1988}). A major limitation of blocking, similar to that of deterministic linking, is that variables used to block records may have different values across files (\cite{Jaro1989}, \cite{Winkler1994}). These differences may place records in the wrong block, resulting in missed true links and possible false links. To address this limitation, blocking methods that partition record pairs into groups based on similarity of the blocking fields rather than exact agreement have been proposed (\cite{Baxter2003}, \cite{Christen2012}, \cite{Steorts2014}). These methods prioritize reducing the dimension of the linkage problem while maintaining a high proportion of true links that are identified. These blocking procedures are sensitive to errors among the blocking fields, and they do not adjust for potential errors in the blocking process in a subsequent analysis.

\citet{Dalzell2018} extended the work of \citet{Gutman2013} to account for possible errors within blocking variables. The extended method assumes that the values of blocking fields may be recorded with errors, and it allows for individual records to shift between blocks. Multiple blocking variable values and linkage structures are sampled from the posterior distribution in order to propagate the errors in the blocking and linking processes. 

All of these blocking methods assume that the blocking variables are recorded in both files. However, they do not address cases where the blocking variable represent similar entities in both files, but unique identifiers are not available to match blocks across files. Specifically, in healthcare settings, patients receive care according to providers (hospitals, nursing-homes, physicians, etc.). These providers create a natural blocking scheme. However, because of confidentiality restrictions, providers' identifiers are unique to each file. Thus, patients receiving treatment from the same provider can be identified within each file separately, but the providers cannot be easily linked across files. In scenarios with limited discriminating information, leveraging the information that entities received care from similar providers can improve linkage accuracy.

\begin{comment}
\citet{Dalzell2018} proposes an extension of \citet{Gutman2013} that accounts for possible errors within blocking variables by re-sampling the values of blocking fields for records, which enables individual records to shift the block it belongs to between iterations of the MCMC algorithm. Multiple blocking variable values and linkage structures are generated to propagate the errors in the blocking and linking process in the subsequent analysis. However, this method assumes that the way by which records are partitioned into groups according to their blocking fields is not known and needs to be estimated. Furthermore, block overlap information is included among the previously mentioned methods as either a minimum block size requirement to ensure privacy protection, or as frequency weights to differentiate informativeness for certain blocking fields. To our knowledge, a method that incorporates expected block overlap size within the likelihood of the linkage structure does not exist.
and does not handle the scenario where the partitioning structure is known, but the labels facilitating how the blocks should pair with each other is missing and needs to be estimated. Furthermore, block overlap information is included among the previously mentioned methods as either a minimum block size requirement to ensure privacy protection, or as frequency weights to differentiate informativeness for certain blocking fields. To our knowledge, a method that incorporates expected block overlap size in the estimation of the blocking structure does not exist.
\end{comment}

We propose a Bayesian multi-layer record linkage method that identifies providers and the corresponding entities treated within each provider. This method has improved linkage accuracy when providers' and individuals' linking variables are limited in their quality of information or are prone to error. The algorithm generates multiple linked datasets, which propagates the uncertainty generated from the linking processes, resulting in valid statistical inferences. We implement the proposed method to examine the associations between functional status assessments and TBI patients' ability to be independent following their injury by linking records from Medicare enrollment records and claims data provided by the Centers for Medicare and Medicaid Services to patient records in the National Trauma Data Bank.

\section{Methods}
\subsection{Data Structure and Notation}
Let $\mathbf{F}_1$ and $\mathbf{F}_2$ be two files comprising $n_1$ and $n_2$ records, respectively. We assume that records $f_{1i} \in \mathbf{F}_1,$ $i=1, \dots, n_1$ are partitioned into $s = 1, \dots, S$ blocks, such that block $s$ comprise $n_{1s}$ records and $\sum_{s=1}^{S} n_{1s}=n_1$. Similarly, $f_{2j} \in \mathbf{F}_2$, $j=1, \dots, n_2$ are partitioned into $t=1, \dots, T$ blocks, where block $t$ comprise $n_{2t}$ records and $\sum_{t=1}^T n_{2t}=n_2$. We assume that the blocks are observed in each file, but the labels linking the blocks across the two files are missing. Let $\mathbf{B}=\{B^{st} \}$ be a latent $S \times T$ binary matrix where
\begin{equation}
B^{st}=
     \begin{cases} 
     1 & \text{if blocks $s \in \mathbf{F}_1$ and $t \in \mathbf{F}_2$ are the same partition.} \\
     0 & \text{Otherwise.}
     \end{cases}
\end{equation}
Without a loss of generality, assume that $S\leq T$, and that the blocks in $\mathbf{F}_1$ will have exactly one match with blocks in $\mathbf{F}_2$. Throughout, this blocking structure will be referred to as complete one-to-one blocking. To enforce this structure, we place the following constraints on $\mathbf{B}$: $\sum_s B^{st}=1, \sum_t B^{st} \leq 1, \sum_s \sum_t B^{st}=S$.

Among block pairs that represent the same entity, the linkage of records within each partition is estimated. Let $\mathbf{C}^{st}$ be a latent $n_{1s} \times n_{2t}$ binary matrix that represents the record-level linkage structure in block pair $\{(s,t) , s \in \mathbf{F}_1, t \in \mathbf{F}_2\}$ such that the $(i,j)^{th}$ entry is
\begin{equation}
C_{ij}^{st}=
     \begin{cases} 
     1 & \text{if record $i$ from block $s \in \mathbf{F}_1$ is linked with record $j$ from block $t \in \mathbf{F}_2$} \\
     0 & \text{Otherwise.}
     \end{cases}
\end{equation}
The sub-matrix $\mathbf{C}^{st}$ is a non-zero matrix when blocks $s$ and $t$ are linked together $(B^{st}=1)$, and is a zero matrix for all other block pairs. To enforce one-to-one record-level linkage within each block pair, the following constraints are applied: $\sum_{i^s} C_{ij}^{st} \leq 1, \sum_{j^t} C_{ij}^{st} \leq 1, \sum_{i^s} \sum_{j^t} C_{ij}^{st} \leq \min(n_{1s}, n_{2t})$. Combining these constraints with the constraints on $\mathbf{B}$ ensures that the record-level linkage structure is one-to-one.

Record $i^s \in \mathbf{F}_1$ comprises of block level and record level variables, $\mathbf{X}_{i}^s= \{ \mathbf{X}_{B}^s, \mathbf{X}_{Ci}^s\}$, where $\mathbf{X}_{B}^s = \{X_{B1}^s, \dots, X_{BP}^s \}$ represents the set of $P$ block-level variables, which are the same for all units in block $s$, and $\mathbf{X}_{Ci}^s=\{X_{Ci1}^s, \dots, X_{CiK}^s \}$ denotes the set of K record-level variables. Similarly, record $j^t \in \mathbf{F}_2$ comprises of $\mathbf{X}_{j}^t=\{\mathbf{X}_{B}^t, \mathbf{X}_{Cj}^t \}$, where $\mathbf{X}_{B}^t = \{X_{B1}^t, \dots, X_{BP}^t \}$ and $\mathbf{X}_{Cj}^t = \{X_{Cj1}^t, \dots, X_{CjK}^t \}$ are the block-level and record-level identifying information, respectively. Let $\mathbf{X}^s=\{\mathbf{X}_i^s \}$ for $i^s=1, \dots, n_{1s}$ and $\mathbf{X}^t=\{ \mathbf{X}_j^t\}$ for $j^t=1, \dots, n_{1t}$ be the collection of blocking and linking variables for blocks $s \in \mathbf{F}_1$ and $t \in \mathbf{F}_2$, respectively. Further, let $\mathbf{X}_1=\{ \mathbf{X}^s\}$ for $s=1, \dots, S$ denote the collection of information in $\mathbf{F}_1$ and $\mathbf{X}_2=\{ \mathbf{X}^t\}$ for $t=1, \dots, T$ denote the information present in $\mathbf{F}_2$.

For each pair of records $i^s \in \mathbf{F}_1$ and $j^t \in \mathbf{F}_2$, let $\mathbf{\Gamma}_{ij}^{st}=\{\mathbf{\Gamma}_{B}^{st}, \mathbf{\Gamma}_{Cij}^{st} \}$ be an agreement vector on the $P+K$ identifying variables such that $\mathbf{\Gamma}_B^{st}=\{\Gamma_{Bp}^{st}:p=1, \dots, P\}$ and $\mathbf{\Gamma}_{Cij}^{st}=\{ \Gamma_{Cijk}^{st}: k=1, \ldots, K\}$. The agreement vector is calculated based on a deterministic function that evaluates the similarity of the two values (\cite{Winkler1989}, \cite{Winkler2006}). The simplest deterministic function examines whether two variables have the same value (\cite{Fellegi1969}). Formally, block-level comparisons are defined as 
\begin{equation}
\Gamma_{Bp}^{st}=
     \begin{cases} 
     1 & \text{if $X_{Bp}^s=X_{Bp}^t$} \\
     0 & \text{Otherwise}
     \end{cases} \label{3}
\end{equation}
for $p=1, \dots, P$, and record-level comparisons as
\begin{equation}
\Gamma_{Cijk}^{st}=
     \begin{cases} 
     1 & \text{if $X_{Cik}^s=X_{Cjk}^t$}. \\
     0 & \text{Otherwise}
     \end{cases}  \label{4}
\end{equation}
for $k=1, \dots, K$.

\subsection{Multi-Layer Mixture Model to Estimate Block and Linkage Structure}
\begin{singlespace}
\begin{figure}
\begin{tikzpicture}[level distance=2cm,
level 1/.style={sibling distance=7cm},
level 2/.style={sibling distance=3.5cm},
level 3/.style={sibling distance=2cm},
level 4/.style={sibling distance=0.25cm},
tree node/.style={circle,draw},
every child node/.style={tree node}]
\node[tree node] (Root) [] {$\mathbf{F}_1 \times \mathbf{F}_2$}
    child{
    node {$B_M$} 
    child { node{}
        child{ node{$C_M$} child{} child{} child{} child{}}
        child{ node{$C_U$} child{} child{} child{} child{}}}
    child { node {}
        child{ node{$C_M$} child{} child{} child{} child{}}
        child{ node{$C_U$} child{} child{} child{} child{}}}
}
child {
    node {$B_U$}
    child { node {} 
    child { node{$C_{NB}$} child{} child{} child{} child{}} }
    child { node {} 
    child { node{$C_{NB}$} child{} child{} child{} child{}}}
};

\node[align=center, text width=3cm] at (-8.25,-2.75) {Block-Level};
\node[align=center, text width=3cm] at (-8.25,-7) {Record-Level};
\node[align=center, text width=1.5cm] at (-5, -2) {\footnotesize True Blocks};
\node[align=center, text width=1.5cm] at (5, -2) {\footnotesize False Blocks};
\draw[dashed] (-8,-1) -- (7,-1);
\draw[dashed] (-8,-4.75) -- (7,-4.75);
\node[align=center, text width=1.5cm] at (-6.25, -8.5) {\footnotesize True Links};
\node[align=center, text width=2cm] at (-4.25, -8.75) {\footnotesize Non-Links within True Blocks};
\node[align=center, text width=3cm] at (3.5, -8.75) {\footnotesize Non-Links within False Blocks};
\end{tikzpicture}
\caption{Graphical Representation of the Multi-Layered Record Linkage Model}
\end{figure}
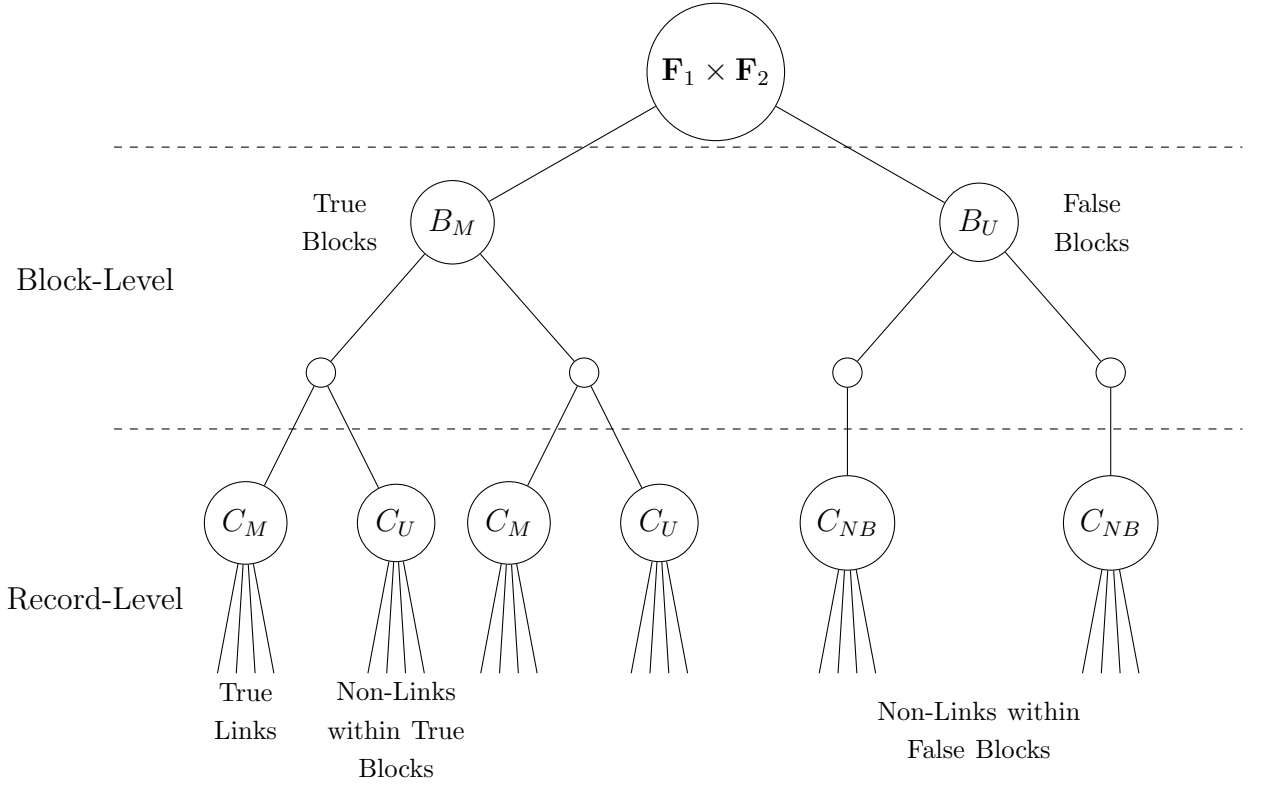

\end{singlespace}

Using the framework of \citet{Fellegi1969}, we consider the set of all pairs of blocks in $\mathbf{F}_1 \times \mathbf{F}_2$ as the union of two disjoint sets of true blocks $\mathbf{B_M}=\{(s,t): s \in \mathbf{F}_1, t \in \mathbf{F}_2, B^{st}=1 \}$ and non-blocks $\mathbf{B_U}=\{(s,t): s \in \mathbf{F}_1, t \in \mathbf{F}_2, B^{st}=0 \}$. For each block pair $(s,t) \in \mathbf{B_M}$, we consider two disjoint sets of individual-level record pairs: true links, $\mathbf{C_M}=\{(i^s, j^t): i^s \in s, s \in \mathbf{F}_1, j^t \in t, t \in \mathbf{F}_2, C_{ij}^{st}=1, B^{st}=1 \}$, and non-links, $\mathbf{C_U}=\{(i^s, j^t): i^s \in s, s \in \mathbf{F}_1, j^t \in t, t \in \mathbf{F}_2, C_{ij}^{st}=0, B^{st}=1 \}$. All record pairs within $\mathbf{B_U}$ belong to the set $\mathbf{C}_{NB}= \{(i^s, j^t): i^s \in s, s \in \mathbf{F}_1, j^t \in t, t \in \mathbf{F}_2, C_{ij}^{st}=0, B^{st}=0 \}$. A graphical representation of this multi-layer linkage structure is depicted in Figure 1.

Let $\Theta=(\theta_{BM},\theta_{BU},\theta_{CM},\theta_{CU}, \theta_{CNB})$ represent the parameters governing the distributions of $\{\mathbf{\Gamma}_{ij}^{st}\}$. We model the distributions of $\mathbf{\Gamma}_{ij}^{st}$ using the following mixture model:
\begin{subequations}
\begin{align}
    \mathbf{\Gamma}_{ij}^{st}|B^{st}=1, C_{ij}^{st}=1 & \sim f(\mathbf{\Gamma}_{B}^{st}|\theta_{BM}) f(\mathbf{\Gamma}_{Cij}^{st}|\theta_{CM}) \label{5a}\\
    \mathbf{\Gamma}_{ij}^{st}|B^{st}=1, C_{ij}^{st}=0 & \sim f(\mathbf{\Gamma}_{B}^{st}|\theta_{BM})f(\mathbf{\Gamma}_{Cij}^{st}|\theta_{CU}) \label{5b}\\
    \mathbf{\Gamma}_{ij}^{st}|B^{st}=0, C_{ij}^{st}=0 & \sim f(\mathbf{\Gamma}_{B}^{st}|\theta_{BU})f(\mathbf{\Gamma}_{Cij}^{st}|\theta_{CNB}), \label{5c}
\end{align}
\end{subequations}
\begin{comment}
    \mathbf{B} &\sim \mathcal{U}\bigg(0, \binom{S}{\min(S,T)} \binom{T}{\min(S,T)} \min(S,T)!\bigg) \label{5d}\\
    \mathbf{C}_{st},n_{mst}|B_{st}=1 &\sim p(\mathbf{C}_{st},n_{mst}|B_{st}=1)\label{5e}
\end{comment}
where $\Theta=(\theta_{BM},\theta_{BU},\theta_{CM},\theta_{CU}, \theta_{CNB})$ is a set of unknown parameters and $s\in \{1, \dots, S\}$; $t\in \{1, \dots, T\}$; $i_s\in \{1, \dots, n_{1s}\}$; $j_t \in \{1, \dots, n_{2t}\}$. Equation \ref{5a} represents the distribution of $\Gamma_{ij}$ for true record pairs residing within true block pairs, Equation \ref{5b} models the distribution of $\Gamma_{ij}$ for non-linking record pairs within true blocks, and Equation \ref{5c} models the distribution of $\Gamma_{ij}$ for individual record pairs within non-linked blocks. The joint likelihood for this mixture model can be written as
\begin{align}
    \mathcal{L}(\mathbf{B}, \mathbf{C}, \Theta|\mathbf{X}_1, \mathbf{X}_2) =  \prod_{s=1}^{S} \prod_{t=1}^T & \bigg[ f(\mathbf{\Gamma}_{B}^{st}|\theta_{BM})
    \prod_{i^s=1}^{n1s} \prod_{j^t=1}^{n2t} \bigg\{ f(\mathbf{\Gamma}_{Cij}^{st}|\theta_{CM})\bigg\}^{C_{ij}^{st}} \bigg\{f(\mathbf{\Gamma}_{C ij}^{st}|\theta_{CU} \bigg\}^{1- C_{ij}^{st}} \bigg]^{B^{st}} \nonumber\\
    \times & \bigg[f(\mathbf{\Gamma}_{B}^{st}|\theta_{BU}) \prod_{i^s=1}^{n1s} \prod_{j^t=1}^{n2t} \bigg\{f(\mathbf{\Gamma}_{Cij}^{st}|\theta_{CNB}) \bigg\} \bigg]^{1-B^{st}} \label{6}
\end{align}

In Equation \ref{6}, we assume that the parameters $\theta_{CM}$, $\theta_{CU}$, and $\theta_{CNB}$ do not vary across blocks. When the comparison functions for each blocking and linking variable follow Equations 3 and 4, then $\theta_{BM}=\{\theta_{BMp}:p=1, \ldots, P \}$, $\theta_{BU}=\{\theta_{BUp}:p=1, \ldots, P \}$, $\theta_{CM}=\{\theta_{CMk}:k=1, \ldots, K \}$, $\theta_{CU}=\{\theta_{CUk}:k=1, \ldots, K \}$, and $\theta_{CNB}=\{\theta_{CNBk}:k=1, \ldots, K \}$. A common simplifying assumption made in many record linkage applications is that the components of $\Gamma_{Cij}^{st}$ are conditionally independent given the linkage status of record pairs (\cite{Winkler1989}, \cite{Jaro1989}). Here, we will assume that $\theta_{BMp}$ and $\theta_{BUp}$ are conditionally independent given $\mathbf{B}$, and that $\theta_{CMk}$, $\theta_{CUk}$, and $\theta_{CNBk}$ are conditionally independent given $\mathbf{B}$ and $\mathbf{C}$. Under these conditional independence assumptions, Equation \eqref{6} can be written as
\begin{subequations}
\begin{align}
    \mathcal{L}(\mathbf{B},\mathbf{C},\Theta|\mathbf{X}_1,\mathbf{X}_2) = \prod_{s=1}^S \prod_{t=1}^T &\bigg[ \prod_{p=1}^P \theta_{BMp}^{\Gamma_{Bp}^{st}} (1-\theta_{BMp})^{1-\Gamma_{Bp}^{st}} \label{7a}\\
    \times & \prod_{i^s=1}^{n_{1s}} \prod_{j^t=1}^{n_{2t}} \bigg\{ \prod_{k=1}^K \theta_{CMk}^{\Gamma_{Cijk}^{st}}(1- \theta_{CMk})^{1-\Gamma_{Cijk}^{st}} \bigg\}^{C_{ij}^{st}} \label{7b}\\
    \times &\bigg\{\prod_{k=1}^K \theta_{CUk}^{\Gamma_{Cijk}^{st}} (1-\theta_{CUk})^{1-\Gamma_{Cijk}^{st}} \bigg\}^{1-C_{ij}^{st}}\bigg]^{B^{st}} \label{7c}\\
    \times &\bigg[\prod_{p=1}^P \theta_{BUp}^{\Gamma_{Bp}^{st}}(1-\theta_{BUp})^{1-\Gamma_{Bp}^{st}} \label{7d}\\
    \times & \prod_{i^s=1}^{n_{1s}} \prod_{j^t=1}^{n_{2t}} \bigg\{\prod_{k=1}^K \theta_{CNBk}^{\Gamma_{Cijk}^{st}}(1-\theta_{CNBk})^{1-\Gamma_{Cijk}^{st}} \bigg\}\bigg]^{1-B^{st}} \label{7e}
\end{align}
\end{subequations}
Equation (\ref{7a}) represents the block-level comparisons for true block pairs, Equation (\ref{7b}) denotes the record-level comparisons for true links when $B^{st}=1$, Equation (\ref{7c}) represents the record-level comparisons for non-linking record pairs when $B^{st}=1$, Equation (\ref{7d}) represents the block-level comparisons for non-block pairs, and Equation (\ref{7e}) represents all non-linking record pairs among non-block pairs. 

When all of the comparisons follow Equations (3) and (4), we assume independent prior beta distributions for computational simplicity: $\theta_{BMp} \sim \mathrm{Beta}(\alpha_{BMp}, \beta_{BMp})$, $\theta_{BUp} \sim \mathrm{Beta}(\alpha_{BUp}, \beta_{BUp})$, $\theta_{CMk} \sim \mathrm{Beta}(\alpha_{CMk}, \beta_{CMk})$, $\theta_{CUk} \sim \mathrm{Beta}(\alpha_{CUk}, \beta_{CUk})$, and $\theta_{CNBk} \sim \mathrm{Beta}(\alpha_{CNBk}, \beta_{CNBk})$ for $p=1, \dots, P$ and $k=1, \dots, K$. To complete the Bayesian model, we define the prior distribution $p(\mathbf{B},\mathbf{C})=p(\mathbf{B})p(\mathbf{C}|\mathbf{B})$. First, we assumed a uniform prior distribution over all possible $\frac{\max(S,T)!}{(\max(S,T)-\min(S,T)!}$ $\mathbf{B}$ matrices that fulfill the complete one-to-one blocking constraints. Given the blocking structure $\mathbf{B}$, the prior for $\mathbf{C}^{st}$ when $B^{st}=0$ is a point mass over the zero matrix. For $\mathbf{B}^{st}=1$, we used a prior distribution for $\mathbf{C}^{st}$ proposed by \citet{Sadinle2017}. Define $n_m^{st}$ as the number of true links in block pair $(s,t)$, where $n_m^{st} \leq \min(n_{1s},n_{2t})$. Each $n_m^{st}$ is given an independent $\mathrm{Binomial}(\min(n_{1s},n_{2t}), \pi)$ prior distribution $ \forall (s,t):B^{st}=1$, where $\pi \sim \mathrm{Beta}(\alpha_{\pi}, \beta_{\pi})$ a-priori. Conditional on $n_m^{st}$, the prior for $\mathbf{C}^{st}$ is uniform over all $\binom{n_{1s}}{n_m^{st}} \binom{n_{2t}}{n_m^{st}}n_m^{st}!$ matrices that satisfy one-to-one linking with $n_m^{st}$ links. The probability mass function for $\mathbf{C}^{st}$ marginalized over $\pi$ can be written as
\begin{align}
    &p(\mathbf{C}^{st},n_{m}^{st}|B^{st}=1,\alpha_{\pi},\beta_{\pi})= \nonumber\\
    &\dfrac{(\max(n_{1s},n_{2t})-n_{m}^{st})!}{\max(n_{1s},n_{2t})!} \dfrac{\Gamma(\alpha_{\pi}+\beta_{\pi})}{\Gamma(\alpha_{\pi})\Gamma(\beta_{\pi})}\dfrac{\Gamma(n_{m}^{st}+\alpha_{\pi})\Gamma(\min(n_{1s},n_{2t})-n_{m}^{st}+\beta_{\pi})} {\Gamma(\min(n_{1s},n_{2t})+\alpha_{\pi}+\beta_{\pi})}. \label{8}
\end{align}
where $\Gamma(.)$ represents the Gamma function (\cite{Sadinle2017}).

\begin{comment}
\begin{equation}
    p(\mathbf{C}^{st},n_{m}^{st}|B^{st}=1,\alpha_{\pi},\beta_{\pi})= \dfrac{(\max(n_{1s},n_{2t})-n_{mst})!}{\max(n_{1s},n_{2t})!} \dfrac{\mathrm{B}(n_m^{st}+\alpha_{\pi}, min(n_{1s},n_{2t})-n_m^{st}+\beta_{\pi})}{\mathrm{B}(\alpha_{\pi},\beta_{\pi})}. 
\end{equation}
\end{comment}

Estimating the linkage structure in Equation (7) using either maximum likelihood methods or by sampling from a Bayesian posterior distribution can be computationally intensive. One possible approximation algorithm would be to first estimate $\mathbf{B}$ using only \eqref{7a} and \eqref{7d}. Using the sampled blocking structures, record-level links are then sampled within linked blocks. We will refer to this approach as Conditionally Independent Bayesian Record Linkage (CIBRL). This approximation ignores the information in the conditional distribution of record-level links when estimating $\mathbf{B}$. For a mixture of two Normal distributions, a larger gain in information is observed as the separation between the two distributions increases (\cite{Titterington1985}). We would expect $\theta_{CM}$ to have larger separation from $\theta_{CU}$ within true block pairs than within false blocks. This larger separation should provide more efficient estimation of the block-level linkage and the record-level linkage when estimated jointly. We propose a MCMC algorithm that jointly samples configurations of $\mathbf{B}$ and $\mathbf{C}$ to improve the accuracy of both the block-level pairings and the record-level linkage, which we will refer to as Multi-Layered Bayesian Record Linkage (MLBRL).

\subsection{Gibbs Sampling Algorithm}
To estimate the linkage structure that preserves the hierarchical partitioning between blocked units, we propose the following MCMC algorithm to obtain posterior samples of $(\mathbf{B},\mathbf{C})$. Directly sampling from the joint posterior of $(\mathbf{B},\mathbf{C})$ would require computing the likelihood for all $\binom{T}{S}S!$ blocking permutations multiplied by all possible $\binom{n_{1s}}{n_{m}^{st}}\binom{n_{2t}}{n_{m}^{st}}n_{m}^{st}!$ linking configurations within each true block pair. This is computationally demanding, even with small number of records and partitions. To address this limitation, we adopt a version of the Metropolis Hastings algorithm that proposes local updates to the status of each block pair $B^{st}$ (\cite{Wu1995}) and a Gibbs sampling algorithm to iteratively update the linking configuration $\mathbf{C}^{st}$ within each true block pair (\cite{Sadinle2017}). 
Starting with random assignments for the block pairs in $\mathbf{B}$ and the record pairs in $\mathbf{C}$, our MCMC sampling algorithm iterates through the following steps:

\begin{enumerate}
    \item Sample new values of $\theta_{BMp}^{[v+1]},\theta_{BUp}^{[v+1]},\theta_{CMk}^{[v+1]},\theta_{CUk}^{[v+1]},\theta_{CNBk}^{[v+1]}$, for $k=1, \dots, K$ and $p=1, \dots, P$, from their respective posterior distributions.
    \item For each true block pair $\mathbf{B}^{[v]*}=\{\mathbf{B}^{[v]}:B^{st[v]}=1, \forall s=1, \dots, S; t=1, \dots, T \}$, sample new values of $\mathbf{C}^{st[v+1]}$ and $n_{m}^{st[v+1]}$. Details are provided in Appendix A.
    \item Let block $s \in \mathbf{F}_1$ be paired with block $t \in \mathbf{F}_2$ at iteration $[v]$. Randomly propose a new block $r \in \mathbf{F}_2$ to form a pair with block $s$. Depending on the blocking status of $r$, there are two possible updates:
    \begin{enumerate}
    \item {If block $r$ does not form a true block pair with any block in $\mathbf{F}_1$ at iteration $[v]$, propose the new true block pair $B^{sr}=1$ and $B^{st}=0$.}
    \item {If block $r$ forms a true block pair with $q \in \mathbf{F}_1$ at iteration $[v]$, propose swapping block designations to form the true block pairs $B^{sr}=1$ and $B^{qt}=1$.}
    \end{enumerate}
    Details of the Metropolis Hastings algorithm for these updates are provided in Appendix B.

\end{enumerate}
At each iteration, this algorithm updates the linkage configuration for each $\mathbf{C}^{st}$ in linked block pairs, and the blocks' pairing. Because the block-level and record-level linking dimensions remain the same at every iteration, we do not need to introduce latent terms, as in a reversible jump MCMC algorithm (\cite{Green1995}).

Linking two files is not commonly the goal of many studies; instead, researchers seek to estimate associations between variables that exist within the merged dataset. The proposed linking algorithm is computationally intensive. Moreover, performing both record linkage and the subsequent analysis within the Bayesian framework is even more computationally intensive. One possible solution is to use a multiple imputation algorithm to average over estimates of the linkage structures (\cite{Gutman2013}, \cite{Sadinle2018}). Specifically, we generate multiple linked datasets using the proposed procedure, analysis is performed separately on each dataset, and overall point and interval estimates are obtained through common combining rules (\cite{Rubin1987}). This procedure properly propagates the error in the linkage while reducing the computation. Formally, $M$ posterior samples of $\mathbf{C}$ are obtained to represent $M$ different linked datasets. Let $\hat{Q}^{(m)}$ represent the desired quantity of interest estimated using the $m^{th}$ linked dataset, and let $U^{(m)}$ represent its estimated sampling variance. The multiple imputation point estimate for $Q$ is $\hat{Q}=m^{-1}\sum_{m=1}^M \hat{Q}^{(m)}$, and its estimated sampling variance is $T=\bar{U}+(1+m^{-1})B$ where $\bar{U}=m^{-1}\sum_{m=1}^M U^{(m)}$ and $B=(m-1)^{-1}\sum_{m=1}^M (Q^{(m)}-\bar{Q})^2$. Inference on $\bar{Q}$ is made according to a Student's t approximation $(\bar{Q}-Q)/\sqrt{T} \sim t_{\nu}$ with degrees of freedom $\nu^{-1}=(m-1)^{-1}((1+m^{-1})B/T )^2$ (\cite{Barnard1999}).

\begin{comment}
In the method proposed by \citet{Dalzell2018}, record pairs shift between block pairs at every iteration, which changes the size of blocks and the number of record pair comparisons and necessitates the inclusion of latent terms to preserve dimensions. In this proposed algorithm, the blocking and linking dimensions remain the same at every iteration. Instead, a new block pair is sampled at every iteration from which to calculate the linkage structure within, and a shift between block pairs results in all of the record pairs in the previous block pair being unlinked and an entirely new linking configuration in the proposed block pair estimated in its place. 
\end{comment}

\section{Simulation Studies}
We compare the performance of MLBRL to CIBRL and to the Bayesian record linkage approach developed by \citet{Sadinle2017} (BRL) using simulations. In CIBRL, the blocking structure is estimated using only the block-level linking variables. Within linked blocks, the record-level linkage structure is estimated using record-level linking variables. In BRL, the linkage structure of individual record pairs is estimated using both record-level and block-level linking variables. BRL does not enforce restrictions between block pairs and allows records from a block $s$ in file $\mathbf{F}_1$ to link with records from multiple blocks in file $\mathbf{F}_2$.

We examine the performance of each method in identifying true block pairs and true record pairs under combinations of high (40\%), medium (20\%), and no error (0\%) among block-level and record-level linking variables. We consider a simulation setting with $S=30$ blocks in $\mathbf{F}_1$ and $T=40$ blocks in $\mathbf{F}_2$. There are $n_{1s}=20$ records within each block $s \in \mathbf{F}_1$, and $n_{2t}=30$ records within each block $t \in \mathbf{F}_2$, such that each true block pair $(s,t) \in B_M$ contains $n_{m}^{st}=15$ true links. Three weakly informative blocking variables are simulated to represent Hospital Region, Hospital Status (public or private), and hospital Trauma Level (I or II), as well as an informative blocking variable that represents the median household Income of the hospital region. In addition, Gender and a continuous age variable are created for each individual record in both files as record-level linking variables. Data is generated for true block pairs and true linking record pairs in $\mathbf{F}_1$, and the values are replicated in $\mathbf{F}_2$. Values for block-level and record-level variables for non-blocks or non-links are simulated from the same distributions as true blocks and true links, but they are not replicated. Table 1 depicts the simulated block-level and record-level linking variables.

\begin{table}[]
\caption{Simulated Block-Level and Record-Level Variables}
\centering
\begin{tabular}{llll}
\hline
Field & Variable Type & Distribution & Levels \\ \hline
Region & Blocking & Discrete Uniform & 4 \\
Hospital Status & Blocking & Bernoulli(.8) & 2 \\
Trauma Level & Blocking & Bernoulli(.5) & 2 \\
Area Income & Blocking & N(50,000, 10,000) & Continuous \\
Date of Birth & Linking & Age $\sim$ N(30,4) & Converted to Year and Month \\
Gender & Linking & Bernoulli(.5) & 2 \\ \hline
\end{tabular}
\end{table}

Equation (\ref{3}) is used to assess agreement among the block-level variables Region, Hospital Status, and Trauma Level. Area income is compared by determining whether the absolute difference between two blocks was less than 500. The continuous age values are converted to date of birth (DOB) with a year, month, and day of birth. We performed two simulation configurations for linking DOB. Day of birth is included in the linkage process in one configuration, while it is withheld in the other to mimic confidentiality restrictions. We present the results when day of birth is withheld in this section, and provide results when it is included in Appendix D.  

Three levels of similarity are used to compare the components of DOB: agreement on DOB year and month is considered the highest level of agreement $(l=3)$, followed by agreement on DOB year only $(l=2)$, and finally no agreement on DOB year $(l=1)$ (\cite{Tromp2006}). The likelihood for $\Gamma_{ijDOB}^{st}$ is $\{\prod_{l=1}^3 \theta_{CMDOB}^{\mathbbm{1}(\Gamma_{Cij}^{st}=l)} \}^{C_{ij}^{st} B^{st}}$ for $(i^s, j^t) \in \mathbf{C}_M$, $\{\prod_{l=1}^3 \theta_{CUDOB}^{\mathbbm{1}(\Gamma_{Cij}^{st}=l)} \}^{(1-C_{ij}^{st}) B^{st}}$ for $(i^s, j^t) \in \mathbf{C}_U$, and $\{\prod_{l=1}^3 \theta_{CNBDOB}^{\mathbbm{1}(\Gamma_{Cij}^{st}=l)} \}^{1-B^{st}}$ for $(i^s, j^t) \in \mathbf{C}_{NB}$, where $\mathbbm{1}{\cdot}$ is an indicator function that is equal to 1 when the condition is fulfilled and 0 otherwise. The remaining set of blocking and linking variables are modeled according to the joint likelihood specified in Equation (7), with each agreement among each blocking and linking variable assumed to be conditionally independent given the blocking and linking structures, respectively. Dirichlet$(1, \dots, 1)$ prior distributions are used for $\theta_{CMDOB}$, $\theta_{CUDOB}$, and $\theta_{CNBDOB}$, and Beta(1,1) distributions are used for the remaining parameters governing the block-level and record-level variables. We also assume $n_m^{st} \sim \mathrm{Beta}(1,1), \forall (s,t) \in \mathbf{B}_m$.

Varying degrees of error among the block-level variables Region and Area Income, as well as the record-level variable DOB month, for records in $\mathbf{F}_1$ were examined.  For a given error rate $\epsilon=(0.0, 0.2, 0.4)$, a block in $\mathbf{F}_1$ would have their true Region value randomly re-sampled with probability equal to $\epsilon$. Errors in Area Income are generated by perturbing each Area Income entry in $\mathbf{F}_1$ with noise that is distributed $N(0, \frac{500}{\Phi^{-1}(1-\epsilon/2)})$. This implies that the probability of categorizing an Area Income comparison as disagreement for true block pairs is equal to $\epsilon$. The DOB month for records in $\mathbf{F}_1$ with true links were randomly altered to a different month with probability $\epsilon$. Errors were introduced into each block-level and record-level variable independently. A total of 100 simulated sets of data were generated for each blocking and linking error configuration.

For each simulated dataset and combination of error rates, we generated 2,000 samples using MCMC algorithms and discarded the first 1,000 samples for all three linkage algorithms. To obtain a linking configuration within true blocks, Step 2 of our proposed algorithm was implemented with 25 iterations per true block pair. We assess the block linking performance of MLBRL and CIBRL by calculating the accuracy $ACC=\frac{\sum_{s=1}^S \sum_{t=1}^T \mathbbm{1}(B^{st}=1 | (s,t) \in \mathbf{B}_M)}{S}$ within each MCMC iteration. The $ACC$ represents the proportion of blocks in $\mathbf{F}_1$ that are linked with their true block pair in $\mathbf{F}_2$. The sensitivity (TPR, or recall), positive predictive value (PPV, or precision), and the F1 score are used to assess the individual record linking performance of MLBRL, CIBRL, and BRL within each MCMC iteration. The sensitivity is calculated as the proportion of true links that are correctly identified, $TPR=\frac{\sum_{s=1}^S \sum_{t=1}^T \sum_{i^s}^{n_{1s}} \sum_{j^t}^{n_{2t}} \mathbbm{1}(C_{ij}^{st}=1|(i^s,j^t) \in \mathbf{C}_M)}{\sum_{s=1}^S \sum_{t=1}^T \sum_{i^s}^{n_{1s}} \sum_{j^t}^{n_{2t}} \mathbbm{1}((i^s,j^t)\in \mathbf{C}_M)}$, and the positive predictive value is the proportion of linked records that are true links, $PPV=\frac{\sum_{s=1}^S \sum_{t=1}^T \sum_{i^s}^{n_{1s}} \sum_{j^t}^{n_{2t}} \mathbbm{1}(C_{ij}^{st}=1|(i^s,j^t) \in \mathbf{C}_M)}{\sum_{s=1}^S \sum_{t=1}^T \sum_{i^s}^{n_{1s}} \sum_{j^t}^{n_{2t}} C_{ij}^{st}}$. The F1 score is equal to $2 \cdot \frac{TPR \cdot PPV}{TPR+PPV}$. For each error configuration and each linkage procedure, we recorded the average of these values across the MCMC iterations and simulated datasets: $\overline{ACC}$, $\overline{TPR}$, $\overline{PPV}$, $\overline{F1}$.

\subsection{Simulation Results}
Table 2 displays the $\overline{TPR}$ and $\overline{PPV}$ of the three methods over the 100 simulated datasets for different combinations of error rates, and Table 3 displays the $\overline{F1}$ score and $\overline{ACC}$. Generally, both MLBRL and CIBRL outperform BRL in terms of $\overline{TPR}$, $\overline{PPV}$, and $\overline{F1}$ score across all error configurations. In configurations with no errors in the block-level or record-level variables, MLBRL and CIBRL perform similarly in terms of $\overline{TPR}$, $\overline{PPV}$, and $\overline{F1}$ score. While the $\overline{ACC}$ is 1 in scenarios without error, the $\overline{TPR}$ less than 1 suggests that the record-level information may not be sufficiently informative to uniquely identify all true links between the data sources. As errors among the block-level variables increases while maintaining no error in the record-level variable DOB, the $\overline{ACC}$ of MLBRL remains nominal at 1 while the $\overline{ACC}$ of CIBRL declines. This decline in average $\overline{ACC}$ results in a large number of true links in blocks that are incorrectly linked being omitted during the record-level linkage, which is reflected in the lower $\overline{TPR}$ and $\overline{F1}$. This implies CIBRL is sensitive to erroneous block-level variables whereas MLBRL can gain additional accuracy from the information contained in the record-level linkage.

As error is introduced to the record-level variable DOB, the $\overline{TPR}$, $\overline{PPV}$, and $\overline{F1}$ score of all three methods decrease. This drop in performance is most severe in BRL. For given error rates on block-level variables, the $\overline{ACC}$ of MLBRL decreases as error among the record-level variable DOB increases, while the $\overline{ACC}$ of CIBRL is independent of errors in the record-level variable. In scenarios when there is no error in the block-level variable Income and there is 40\% error in record-level variable DOB, CIBRL has higher $\overline{ACC}$ than MLBRL. As a result, CIBRL also outperforms MLBRL in terms of $\overline{TPR}$ and $\overline{F1}$ in these scenarios. To examine this scenario further, we examined the simulations in which day of birth is included as a a record-level linking variable. When day of birth is included and with no errors in the blocking variables, the $\overline{ACC}$ was similar for MLBRL and CIBRL (Tables 8 and 9 in Appendix D). This suggests that when the information among record-level linking variables is weak (e.g. large errors among identifying variables), and there are no errors among blocking variables, it may be advantageous to separate the block-level and record-level linkage to prevent the record-level errors from negatively impacting the block-level linkage. However, with the introduction of errors in the blocking variables, MLBRL performs significantly better than CIBRL even in scenarios with large errors in the record-level linking variables.

Figure 2 shows the average log-probability matrix among the true block pairs and true link record pairs using MLBRL, CIBRL, and BRL across 100 datasets that includes day of birth as a linking variable and without error there is 40\% error among Region, Area Income, and DOB month. In an ideal scenario, we would expect the diagonal values to be close 0, which would indicate probabilities close to 1 for the correct classification of true linking block and record-level pairs. Figure 2a shows that the block-level classification when using CIBRL can be inaccurate as some of the off-diagonal block pairs also have log-probability values that are close to 0, whereas MLBRL is robust to errors in the block-level variables (Figure 2b). Because BRL does not constrain the record-level linkage to blocks, it results in records within one block being linked to records from multiple blocks in the other file (Figure 2c). This indicates poor record-level linkage when the error rate is high. Figure 2d shows that CIBRL is able to estimate the record-level linkage well within block-pairs that are correctly identified, but the record-level estimation suffers in blocks with significant uncertainty that are incorrectly paired. MLBRL estimates the blocking structure well even with high block-level errors, and it estimates the record-level linkage accurately (Figure 2e).

\begin{figure}
  \begin{subfigure}[t]{.45\textwidth}
    \includegraphics[width=\linewidth]{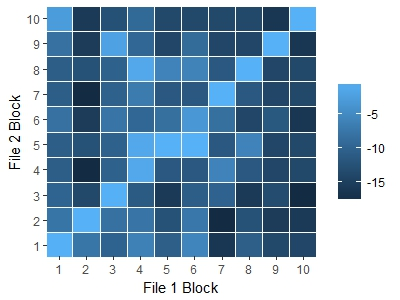}
    \caption{Block Pair Log Probability Density Using CIBRL.}
  \end{subfigure}
  \hfill
  \begin{subfigure}[t]{.45\textwidth}
    \includegraphics[width=\linewidth]{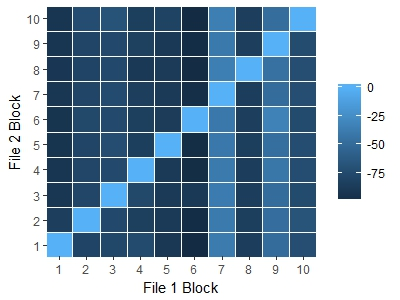}
    \caption{Block Pair Log Probability Density Using MLBRL.}
  \end{subfigure}
  \medskip
  \begin{subfigure}[t]{.45\textwidth}
    \includegraphics[width=\linewidth]{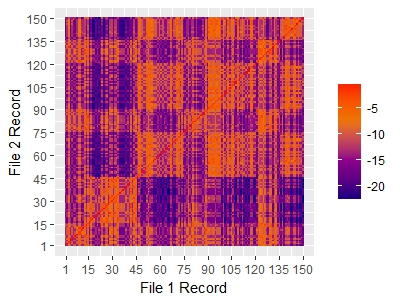}
    \caption{Record Pair Log Probability Density without Blocking Using BRL.}
  \end{subfigure}
  \hfill
  \begin{subfigure}[t]{.45\textwidth}
    \centering
    \includegraphics[width=\linewidth]{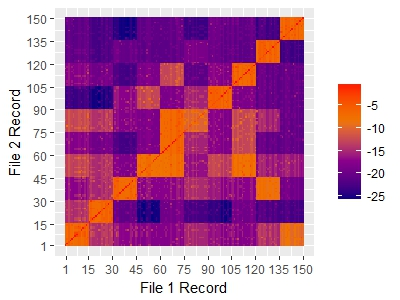}
    \caption{Record Pair Log Probability Density with Blocking Using CIBRL.}
    \end{subfigure}
    \medskip
\begin{subfigure}[t]{.45\textwidth}
    \includegraphics[width=\linewidth]{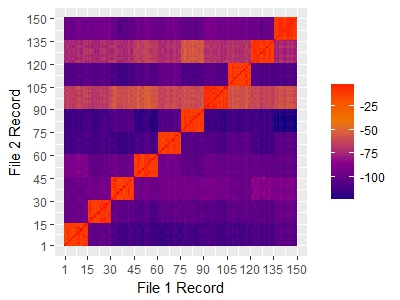}
    \caption{Record Pair Log Probability Density with Blocking Using MLBRL.}
  \end{subfigure}

  \caption{Heatmaps of log-probability matrices for block-level and record-level linkage.}

\end{figure}

\begin{table}[]
\caption{Average TPR and PPV across varying error rates for blocking and linking variables.}
\centering
\begin{tabular}{ccc|ccc|ccl}
 &  &  & \multicolumn{3}{c|}{$\overline{TPR}$} & \multicolumn{3}{c}{$\overline{PPV}$} \\
$\epsilon_{Region}$ & $\epsilon_{Income}$ & $\epsilon_{DOB}$& MLBRL & CIBRL & BRL & MLBRL & CIBRL & BRL \\ \hline
\multirow{9}{*}{0} & \multirow{3}{*}{0} & 0 & 0.88(.014) & 0.86(.026) & 0.73(.017) & 0.80(.026) & 0.81(.030) & 0.59(.020) \\
 &  & 0.2 & 0.69(.018) & 0.68(.024) & 0.60(.019) & 0.70(.029) & 0.70(.027) & 0.46(.019) \\
 &  & 0.4 & 0.49(.023) & 0.54(.025) & 0.41(.022) & 0.58(.030) & 0.58(.030) & 0.34(.021) \\
 & \multirow{3}{*}{0.2} & 0 & 0.88(.013) & 0.83(.033) & 0.74(.018) & 0.82(.026) & 0.79(.029) & 0.60(.021) \\
 &  & 0.2 & 0.70(.017) & 0.66(.029) & 0.55(.019) & 0.70(.025) & 0.70(.033) & 0.46(.021) \\
 &  & 0.4 & 0.50(.025) & 0.52(.031) & 0.39(.021) & 0.57(.030) & 0.58(.028) & 0.33(.020) \\
 & \multirow{3}{*}{0.4} & 0 & 0.88(.014) & 0.75(.048) & 0.73(.017) & 0.81(.029) & 0.79(.027) & 0.59(.019) \\
 &  & 0.2 & 0.65(.018) & 0.59(.041) & 0.53(.019) & 0.68(.034) & 0.67(.034) & 0.44(.020) \\
 &  & 0.4 & 0.49(.021) & 0.47(.034) & 0.38(.023) & 0.56(.031) & 0.57(.034) & 0.32(.020) \\ \hline
\multirow{9}{*}{0.2} & \multirow{3}{*}{0} & 0 & 0.88(.015) & 0.82(.040) & 0.60(.020) & 0.81(.023) & 0.80(.027) & 0.55(.022) \\
 &  & 0.2 & 0.68(.018) & 0.67(.028) & 0.42(.021) & 0.72(.030) & 0.70(.033) & 0.44(.028) \\
 &  & 0.4 & 0.49(.023) & 0.53(.030) & 0.27(.020) & 0.57(.033) & 0.59(.029) & 0.32(.028) \\
 & \multirow{3}{*}{0.2} & 0 & 0.88(.015) & 0.80(.041) & 0.64(.021) & 0.81(.032) & 0.79(.027) & 0.58(.024) \\
 &  & 0.2 & 0.68(.017) & 0.63(.037) & 0.42(.020) & 0.71(.031) & 0.69(.032) & 0.43(.026) \\
 &  & 0.4 & 0.51(.024) & 0.48(.031) & 0.27(.020) & 0.57(.031) & 0.57(.034) & 0.32(.027) \\
 & \multirow{3}{*}{0.4} & 0 & 0.88(.015) & 0.68(.065) & 0.61(.021) & 0.81(.030) & 0.79(.036) & 0.56(.023) \\
 &  & 0.2 & 0.70(.018) & 0.55(.047) & 0.40(.019) & 0.69(.027) & 0.67(.033) & 0.41(.024) \\
 &  & 0.4 & 0.46(.022) & 0.43(.039) & 0.28(.021) & 0.55(.032) & 0.55(.035) & 0.33(.029) \\ \hline
\multirow{9}{*}{0.4} & \multirow{3}{*}{0} & 0 & 0.88(.016) & 0.83(.037) & 0.55(.023) & 0.81(.026) & 0.80(.029) & 0.51(.022) \\
 &  & 0.2 & 0.70(.018) & 0.66(.031) & 0.36(.023) & 0.70(.028) & 0.70(.032) & 0.41(.027) \\
 &  & 0.4 & 0.52(.023) & 0.51(.034) & 0.22(.021) & 0.56(.029) & 0.57(.030) & 0.29(.028) \\
 & \multirow{3}{*}{0.2} & 0 & 0.88(.014) & 0.71(.050) & 0.52(.022) & 0.81(.024) & 0.78(.030) & 0.49(.022) \\
 &  & 0.2 & 0.67(.018) & 0.57(.038) & 0.36(.023) & 0.69(.028) & 0.66(.031) & 0.40(.026) \\
 &  & 0.4 & 0.46(.022) & 0.44(.036) & 0.21(.020) & 0.54(.035) & 0.55(.033) & 0.28(.029) \\
 & \multirow{3}{*}{0.4} & 0 & 0.88(.014) & 0.61(.060) & 0.54(.022) & 0.81(.023) & 0.75(.033) & 0.50(.023) \\
 &  & 0.2 & 0.70(.018) & 0.49(.052) & 0.36(.023) & 0.71(.029) & 0.65(.041) & 0.40(.026) \\
 &  & 0.4 & 0.46(.022) & 0.38(.040) & 0.21(.020) & 0.55(.034) & 0.51(.038) & 0.28(.028) \\ \hline
\end{tabular}
\end{table}

\begin{table}[]
\caption{Average F1 score and Block Accuracy across varying error rates for blocking and linking variables.}
\centering
\begin{tabular}{ccc|ccc|cc}
 &  &  & \multicolumn{3}{c|}{$\overline{F1}$} & \multicolumn{2}{c}{$\overline{ACC}$} \\
$\epsilon_{Region}$ & $\epsilon_{Income}$ & $\epsilon_{DOB}$& MLBRL & CIBRL & BRL & MLBRL & CIBRL \\ \hline
\multirow{9}{*}{0} & \multirow{3}{*}{0} & 0 & 0.84(.017) & 0.83(.023) & 0.68(.018) & 1.00(.000) & 0.98(.017) \\
 &  & 0.2 & 0.69(.020) & 0.69(.021) & 0.52(.019) & 1.00(.000) & 0.98(.017) \\
 &  & 0.4 & 0.53(.022) & 0.56(.022) & 0.37(.021) & 0.92(.004) & 0.98(.017) \\
 & \multirow{3}{*}{0.2} & 0 & 0.85(.017) & 0.81(.024) & 0.68(.019) & 1.00(.000) & 0.94(.033) \\
 &  & 0.2 & 0.70(.017) & 0.68(.025) & 0.52(.020) & 1.00(.000) & 0.94(.033) \\
 &  & 0.4 & 0.53(.021) & 0.55(.023) & 0.37(.020) & 0.92(.014) & 0.94(.033) \\
 & \multirow{3}{*}{0.4} & 0 & 0.85(.019) & 0.77(.032) & 0.68(.018) & 1.00(.000) & 0.84(.051) \\
 &  & 0.2 & 0.67(.021) & 0.63(.031) & 0.52(.020) & 0.94(.000) & 0.84(.051) \\
 &  & 0.4 & 0.52(.021) & 0.51(.029) & 0.37(.020) & 0.86(.008) & 0.84(.051) \\ \hline
\multirow{9}{*}{0.2} & \multirow{3}{*}{0} & 0 & 0.84(.017) & 0.81(.027) & 0.62(.021) & 1.00(.000) & 0.95(.029) \\
 &  & 0.2 & 0.70(.019) & 0.68(.023) & 0.47(.022) & 0.96(.000) & 0.95(.029) \\
 &  & 0.4 & 0.53(.023) & 0.55(.024) & 0.32(.023) & 0.88(.010) & 0.95(.029) \\
 & \multirow{3}{*}{0.2} & 0 & 0.85(.021) & 0.79(.028) & 0.62(.021) & 1.00(.000) & 0.90(.042) \\
 &  & 0.2 & 0.69(.019) & 0.66(.028) & 0.47(.022) & 0.98(.000) & 0.90(.042) \\
 &  & 0.4 & 0.54(.022) & 0.52(.027) & 0.32(.022) & 0.96(.000) & 0.90(.042) \\
 & \multirow{3}{*}{0.4} & 0 & 0.84(.019) & 0.73(.048) & 0.62(.021) & 1.00(.000) & 0.77(.064) \\
 &  & 0.2 & 0.70(.018) & 0.60(.036) & 0.47(.022) & 1.00(.000) & 0.77(.064) \\
 &  & 0.4 & 0.50(.022) & 0.48(.033) & 0.31(.022) & 0.86(.008) & 0.77(.064) \\ \hline
\multirow{9}{*}{0.4} & \multirow{3}{*}{0} & 0 & 0.85(.018) & 0.82(.026) & 0.56(.022) & 1.00(.000) & 0.94(.038) \\
 &  & 0.2 & 0.70(.018) & 0.68(.025) & 0.41(.023) & 1.00(.000) & 0.94(.038) \\
 &  & 0.4 & 0.54(.021) & 0.54(.028) & 0.26(.023) & 0.94(.011) & 0.94(.038) \\
 & \multirow{3}{*}{0.2} & 0 & 0.85(.016) & 0.74(.034) & 0.56(.022) & 1.00(.000) & 0.80(.052) \\
 &  & 0.2 & 0.68(.019) & 0.61(.029) & 0.41(.024) & 0.97(.000) & 0.81(.052) \\
 &  & 0.4 & 0.50(.023) & 0.49(.030) & 0.26(.023) & 0.84(.006) & 0.81(.052) \\
 & \multirow{3}{*}{0.4} & 0 & 0.84(.015) & 0.67(.046) & 0.56(.023) & 1.00(.000) & 0.69(.067) \\
 &  & 0.2 & 0.70(.019) & 0.56(.044) & 0.41(.024) & 1.00(.000) & 0.69(.067) \\
 &  & 0.4 & 0.50(.022) & 0.43(.036) & 0.26(.023) & 0.86(.006) & 0.69(.067) \\ \hline
\end{tabular}
\end{table}

\section{Application to the National Trauma Data Bank}
A large portion of older patients with TBI requiring post-acute care are discharged to skilled nursing facilities (SNFs), which are designed to provide constant nursing care and assistance with daily activities for its residents (\cite{Tepas2013}, \cite{Lueckel2018}). This group often has lower functional status and worse clinical measures than those that are discharged home directly. While the number of older patients with TBI who are discharged to SNFs has steadily increased, a discharge policy that outlines the types of patients with TBI that should be released to SNF does not exist. In order to examine the recovery potential within SNFs, it is necessary to estimate the relationships between a patient's pre-existing and injury-related clinical characteristics and their health outcomes following admission to SNF. To assess these relationships, we link TBI inpatient admissions that occurred between 2011-2015 among patients 65 and older within a well defined sample of burn bed hospitals (\cite{Klein2009}) from two data sources: the National Trauma Data Bank (NTDB), and Medicare enrollment and claims data.

\subsection{Data}
The NTDB represents the largest aggregation of trauma registry data in the United States (\cite{NTDB}) and contains standardized registry information for all trauma-related admissions at over 900 participating facilities. Data is submitted by a trauma registrar and includes details of facility characteristics, individual demographics and comorbid conditions, all ICD-9-CM diagnosis and procedure codes pertaining to the injury, and discharge disposition. Additionally, the registry contains standardized injury severity indices including the Glasgow Coma Score (GCS), which is an assessment of consciousness recorded at emergency department admission, and the head Abbreviated Injury Scale (hAIS), which catalogues the severity of head related trauma injuries. A total of $n_1=20,570$ TBI admissions with subsequent discharge to SNF are observed in the NTDB between 2011 and 2015 for patients 65 years and older. These incidents originate from $S=91$ trauma centers with burn-beds and approximately 73\% of these cases cite Medicare as their primary payment source.

Medicare enrollment and claims data includes information from three sources. The Medicare Master Beneficiary Summary File provides information about an individual's demographic characteristics, Medicare eligibility, chronic health conditions, and healthcare utilization. Medicare Part A claims data contains hospital admission and discharge information, as well as up to 25 ICD-9-CM diagnosis codes and up to 12 procedure codes that are submitted by the hospital billing agency for the TBI hospitalization event. The Medicare enrollment and claims data were linked with a unique identifier to the 2015 American Hospital Association Survey, which characterizes the trauma facility where the TBI incident was treated. A total of $n_2=23,522$ TBI admissions of individuals aged 65 and older that were discharged to SNF are observed between 2011 to 2015 in Medicare data. These admissions take place within $T=94$ trauma facilities with burn beds. 

Table 4 summarizes the blocking and linking variables that exist in each of the data files, as well as variables that will be used in the analysis and appear only in one of the datasets. Details about the creation of blocking and linking variables can be found in Appendix C. The head AIS and the Glasgow coma scores that are only available in the NTDB, which are originally integer values, are dichotomized into mild and severe categories. The MBSF provides the fee-for-service status for each patient, we calculate whether a patient returned to the community 100 days after SNF admission using Medicare claims.

\begin{table}[]
\caption{Variables in the NTDB and Medicare Data Sets}
\resizebox{\textwidth}{!}{%
\begin{tabular}{l|l|l|l}
\hline
 & Variable Name & Type &Values \\ \hline
\multirow{6}{*}{\begin{tabular}[c]{@{}l} Block-level Variables \\in Both Files \end{tabular}} & Hospital Region & BV &Northeast, South, Midwest, West \\
 & Bed Size &BV & 0-200, 201-400, 401-600, 600+ \\
 & Pediatric Beds &BV & 0, 1+ \\
 & Trauma Level &BV & I, II or III \\
 & ICD-9 Trauma Diagnosis &BV & \begin{tabular}[c]{@{}l@{}}801.7, 801.5-801.9, 802.1, 802.3, 802.5, \\   806.0, 806.4, 812.1-812.5, 851.4, 880.0\end{tabular} \\
 & ICD-9 CM Trauma Procedures &BV & 1.20, 1.28, 76.72, 76.74, 76.76, 76.72-76.79 \\ \hline
\multirow{6}{*}{\begin{tabular}[c]{@{}l} Record-level Variables\\ in Both Files \end{tabular}} & Admission Year & LV& 2011, 2012, 2013, 2014, 2015 \\
& Length of Stay &  LV, X&Integer values greater than 0\\
 & Age &  LV, X&Discrete values 65-89, category for 90+ \\
 & Gender & LV, X& Male, Female \\
 & Race &  LV, X&White, not White \\
 & Type and Severity of TBI &  LV, X& Fracture (I, II, or III), Internal (I or II) \\
 & ICD9 Procedures on claim & LV& 1.09, 1.24, 1.25, 1.31, 8.81 \\ 
 & Chronic Condition Indicators & LV, X& AMI, CHF, COPD, Diabetes, Hypertension \\
 \hline
NTDB only & Head AIS score (hAIS) & X& Mild (1-2), Severe (3-5)  \\
& Total Glasgow coma score (tGCS) & X&Severe (3-12), Mild (13-15)\\
 \hline
Medicare only & \begin{tabular}[c]{@{}l} Return to Community\\ after 100 days \end{tabular}& Y& Yes, No \\
\hline
\end{tabular}%
}
\footnotesize{$^1$BV denotes a blocking variable. \\ 
$^2$LV denotes a linking variable. \\ 
$^3$X denotes a covariate used in the analysis model. \\
$^4$Y denotes the outcome in the analysis model. \\}

\end{table}

\subsection{Record Linkage Implementation}
We implement BRL, CIBRL, and MLBRL to link TBI cases between the NTDB and Medicare files. For each of these linking methods, we enforce exact agreement between hospital region at the block-level, and hospital admission year, age, and gender at the record-level. This reduces the total number of record pair comparisons to $992,239$. In CIBRL and MLBRL, we further constrain the dimension of record pairs such that only records within a true block pair can be designated as links within each iteration of the MCMC algorithm.

\begin{comment}
the block-level variable hospital region and record-level variables hospital admission year, age, and gender. This reduces the number of record pair comparisons in BRL to $N_{BRL}=992,239$ record pairs and CIBRL to $N_{CIBRL}=31,838,161$ record pairs. In HBRL, we are only able enforce exact agreement at the block-level for hospital region since the dimension of record pairs would not be preserved within each block pair if exact agreement were enforced between variables at the record-level. This results in $N_{HBRL}=126,215,480$ record pairs that we need to consider. 
\end{comment}

For BRL, the block-level variables in both files, excluding hospital region, are replicated for each record and treated as record-level linking variables, resulting in a total of $K=36$ linking comparisons. We compare values of hospital length of stay between record pairs by defining agreement as $|LOS_{NTDB}-LOS_{Medicare}| \leq 0.25 \times \max(LOS_{NTDB},LOS_{Medicare})$ and disagreement otherwise. Equations \eqref{3} and \eqref{4} are used to compare the values for each of the remaining linking variables across the two files. We used $\mathrm{Beta}(1,1)$ prior distributions for each of the parameters in $\Theta_M$ and $\Theta_U$, as well as for the expected proportion of links $\pi$. BRL generates 2,000 samples for the record-level linkage $\mathbf{C}$. 

A total of $P=22$ block-level and $K=14$ record-level linking variables are used for CIBRL and MLBRL. The same agreement functions and prior distributions described for BRL were used to compare these variables. Both algorithms generated 2,000 samples of $\mathbf{B}$. In CIBRL, record-level link designations are proposed for the entries in the NTDB using 100 updates of the linkage structure for each sample of $\mathbf{B}$. In MLBRL, the linkage structure for true block pairs and potential block updates are estimated using 100 iterations based on the algorithm described in Appendix A.

\subsection{Linkage Results}
Table 5 provides a summary of the linkage performance of BRL, CIBRL, and MLBRL over $m=100$ imputations of the posterior distribution of the  record-level linkage structure. The distribution of linked records using BRL ranges between 5255 and 5618, has a posterior mean of 5,433, and a 95\% credible interval of [5284, 5597]. CIBRL produces an average of 5,518 linked records, but has a range between 2,479 and 7,314 and a 95\% credible interval of [4015, 7021]. MLBRL yields the highest average number of linked records at 9085, has a range between 4636 and 12450 records, and a 95\% credible interval of [5250, 12281].

\begin{table}[h]
\centering
\caption{Comparison of Linkage Results between BRL, CIBRL, and MLBRL. $\sigma$ represents the square root of the total imputation variance across 100 samples.}
\begin{tabular}{lccccc}
Method & Distribution of $n_m$ & $\hat{n}_m$ & 95\% CI  & $\sigma$ \\
\hline
BRL & \begin{minipage}{.28\textwidth}
      \includegraphics[width=\linewidth]{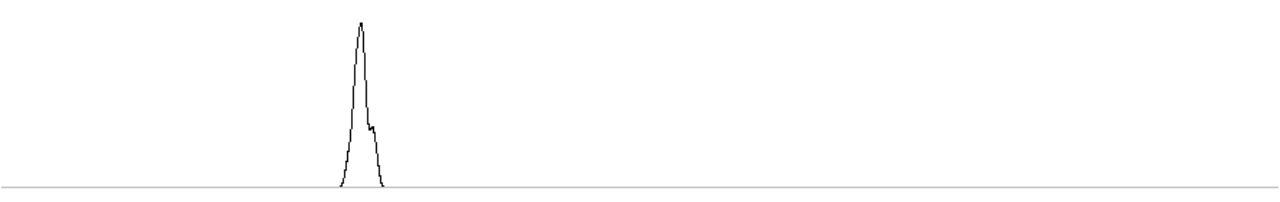}
    \end{minipage} & 5433.2 & (5284.1, 5596.6) & 73.3\\
CIBRL &\begin{minipage}{.28\textwidth}
      \includegraphics[width=\linewidth]{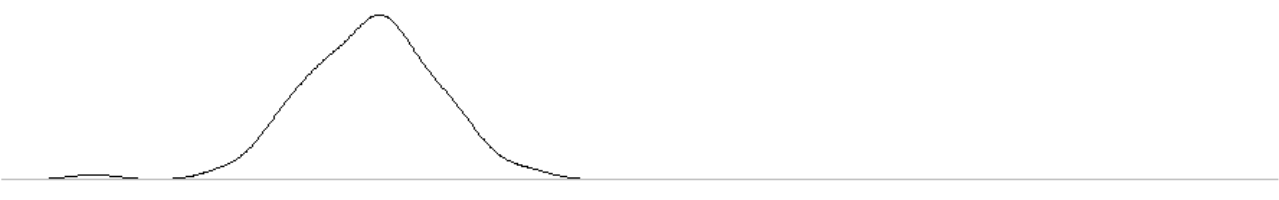}
    \end{minipage}  & 5517.9 & (4140.4, 6813.0) &  757.4\\
MLBRL & \begin{minipage}{.28\textwidth}
      \includegraphics[width=\linewidth]{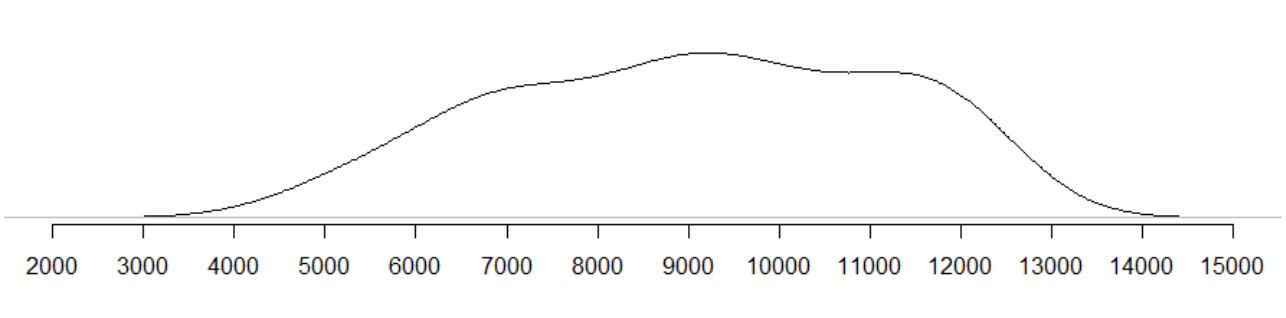}
    \end{minipage} &  9085.4 &  (5250.3, 12281.2) & 2086.2\\
\hline
\end{tabular}
\end{table}

While the point estimate for the number of linked records obtained by BRL and CIBRL is not significantly different, the true link composition given by BRL is different from the links estimated by CIBRL and MLBRL. Under CIBRL and MLBRL, the records within each hospital in the NTDB can only be linked to records from one hospital in the Medicare file at each MCMC iteration. Using BRL, an average of 60.3\% of linked records from hospitals in the NTDB belong to the most frequently paired hospital from the Medicare file (95\% CI: [58.4\%, 62.0\%]). The remaining 39.7\% of linked records comprise an average of 9.52 different hospitals among the posterior samples, with a 95\% credible interval of [9.16, 9.87]. This suggests that BRL may produce linkage estimates with high error levels. 

Compared to BRL and CIBRL, MLBRL yields a higher average number of linked records with a wider credible interval. The incorporation of the patient-level linkage in making linkage decisions between hospitals enables MLBRL to better link hospitals of similar patient sizes. This translates to more true links being identified in blocks that are correctly paired.

\begin{comment}
The BRL performance statistics:\\
Average number of Medicare hospitals within a NTDB hospital partition: 9.52 (9.16, 9.87)\\
Hospital Heterogeneity among the records within a NTDB partition: 60.25\% (58.47\%, 62.02\%).
\end{comment}

\subsection{Analysis Model and Results}
For each imputed complete dataset obtained using BRL, CIBRL, and MLBRL, we form an analytic sample of individuals who are Medicare fee-for-service 100 days following their initial hospital admission. The outcome of interest is whether an individual admitted to SNF following TBI is able to return to the community within 100 days after discharge from a hospital. We examine the association between the outcome and hAIS and tCGS measures separately. To estimate this association, we adjusted for demographic characteristics and pre-existing medical conditions as listed in Table 4. Excluding the variables that are exclusive to the NTDB, we use the Medicare reported values for all of the remaining variables in our analysis. Two logistic regression models were examined for hAIS and tGCS separately:
\begin{equation}
    P(Y=1|\mathbf{X})=\dfrac{exp[(1,\mathbf{X})^T \beta]}{1+exp[(1,\mathbf{X})^T \beta]}. \label{9}
\end{equation}

We calculate point estimates and sampling variance estimates of the marginal odds ratio of high versus low hAIS and GCS score categories using Equation \eqref{9}. Table 6 provides the summaries for the estimated marginal odds ratios and 95\% credible intervals of hAIS and GCS using the three linkage methods. Overall, hAIS and GCS are not found to be significantly associated with successful discharge from SNF after 100 days in any of the three linkage methods. The odds of returning to the community among individuals with high AIS scores (e.g. more severe injury severity) are estimated to be 0.987, 0.993, or 0.990 times less than those with low AIS scores (e.g. more mild injury severity) using linked samples from BRL, CIBRL, and MLBRL respectively. All three methods estimate that the odds of returning to the community are higher for individuals with lower functional status, as recorded by the tGCS. The estimated odds ratio with MLBRL is similar to those obtained by BRL or CIBRL (1.004 vs 1.002 or 1.001). Narrower interval estimates for the odds ratios are obtained by data linked using MLBRL than those obtained using BRL or CIBRL, which is a result of the higher number of linked records.

\begin{table}[h]
\centering
\caption{Comparison of Estimated Odds Ratio of hAIS and GCS between BRL, CIBRL, and MLBRL.}
\begin{tabular}{cccccc}
Variable &Method & OR & 95\% CI  \\
\hline
\multirow{3}{*}{hAIS (Severe vs Mild)}& BRL & 0.987 & (0.941, 1.036)  \\
&CIBRL  & 0.993 & (0.943, 1.046) \\
&MLBRL &  0.990 &  (0.952, 1.029)\\
\hline
\multirow{3}{*}{tGCS (Severe vs Mild)} & BRL & 1.002 & (0.941, 1.068) \\
&CIBRL & 1.001 & (0.934, 1.072) \\
&MLBRL &  1.004&  (0.950, 1.062)\\
\hline
\end{tabular}
\end{table}

\section{Discussion}
Existing Bayesian record linkage methods work well when the linking variables are informative with low error rates. However, their performance can suffer when the linking information is limited with potential errors. These scenarios occur frequently in applications involving large public health datasets. We present a multi-layer Bayesian record linkage approach that simultaneously estimates the linkage of unlabeled block partitions between the two files and the individual records nested within each block. This approach results in higher accuracy when estimating both the record-level and block-level linkages. The algorithm also enables researchers to obtain point estimates and interval estimates that reflect the uncertainty in the block-level and record-level linkages. 

Using simulations, we compare the performance of our newly proposed algorithm to an existing Bayesian record linkage method and an algorithm that considers blocking and linking independently. In simulations, the proposed method generally provides the highest TPR, PPV, F1 score, and block accuracy across different configurations of error rates among the blocking and linking variables. This demonstrates that the record-level linkage can greatly benefit from incorporating blocking structure constraints. Even without any errors in the block-level and record-level variables, MLBRL and CIBRL reduce the number of false links identified compared to BRL. When linking variables are recorded with errors, MLBRL and CIBRL identify more true links compared to BRL. These results appear in the real data analysis where many patients receiving care in the same hospital in one file are linked to patients from multiple hospitals in the other file. CIBRL has lower TPR and F1 score compared to MLBRL when errors exist in the blocking fields. This implies that CIBRL relies heavily on the accuracy of block-level variables, and the assumed independence between the block-level linkage and the record-level linkage may limit this method to scenarios where the block-level variables are recorded with high accuracy. MLBRL correctly identifies the blocking structure with higher accuracy when errors exist among block-level variables. This suggests that incorporating the record-level linkages within block-level linkage results in an algorithm that is more robust to errors among blocking variables. CIBRL performs more favorably with large amounts of error among record-level variables and little error among block-level variables. This is because large error among the record-level variables may negatively impact the relatively accurate block-level linkage when using MLBRL. Commonly in health related datasets, different instruments are used to record providers' characteristics and their values may differ across files. Thus, expecting little error in block-level variables may not be plausible. 

We did not find significant associations between different injury scores and release to the community. Point estimates show that higher tGCS may indicate lower odds of returning to the community after 100 days. While these associations might be contrary to our expectations, the head abbreviated injury severity score and Glasgow coma scores are taken at the beginning of patients' traumatic brain injury at hospital admission. Thus, these may not be reflective of the treatment a patient receives and their recovery during their inpatient stay. Our models show that these scores are not good predictors of recovery among individuals who are assigned to SNF.

While the method described in this paper present a novel approach to increase robustness to record-level linkage, several future extensions can be pursued. In the current approach, we assume that the record-level linking parameters do not vary across blocks. A possible extension is to adopt principles from \citet{Larsen2005} and allow linking variables to vary between blocks using hierarchical prior distributions. Another extension is to incorporate relationships between variables that are exclusive to one file in the block-level linkage and the record level linkage. Lastly, the current algorithm is computationally intensive and development of more computationally efficient algorithms that propagate the errors in the linkage is an area of further research. Possible approaches may rely on the algorithms proposed by \citet{Zanella2019}.

In conclusion, we propose a new record linkage approach that simultaneously estimates block-level and record-level linkage. This new procedure results in improved linkage accuracy and propagates the errors in the linkage status in subsequent analysis through a multiple imputation approach. This algorithm can be used in other settings with limited linking information where there is a known partitioning structure (e.g. individuals within families) within each file that cannot be linked using unique identifiers.

\begin{comment}
\text{Additional Things To Consider:}
\begin{itemize}
    
    \item Talk about how if the blocking information is strong, the linking information does not matter as much. However if the blocking information is weak, then we gain significant information via the linking likelihood.
    \item Perhaps some self-criticism of this method is necessary. We can discuss the computational complexity of this method versus Bayesian record linkage and conditionally independent hierarchical record linkage. The intended use of this method is for situations when the information contained in both the blocking AND the linking variables will give us low precision and recall, so our joint hierarchical framework will significantly improve the accuracy of the linking information. If blocking information is informative but linking information is low, we can use the conditionally independent hierarchical method. If the linking information is strong, we can use standard Bayesian record linkage methods.
\end{itemize}
\end{comment}

\section*{Acknowledgements}
The authors wish to thank Frank DeVone and Emily Evans for their valuable assistance with this project.

\section*{Funding}
This work was supported in part by a grant from the National Institutes of Health under award number R21AG059120. This research was also supported through a Patient-Centered Outcomes Research Institute (PCORI) Award ME-2017C3-9241. Disclaimer: All statements in this report, including its findings and conclusions, are solely those of the authors and do not necessarily represent the views of the PCORI, its Board of Governors or Methodology Committee.

\bibliography{main}

\appendix
\section{Appendix A: Updating the Linkage Structure within True Block Pairs}
In this section, we describe the algorithm that is used to sample the linkage structure $\mathbf{C}^{st}$ within a true block pair $(s,t)$. Given that the linkage structure for record pairs in $(s,t)$ are independent from the linkage structure of other block pairs given $\mathbf{B}$, the linkage likelihood in true block $(s,t)$ given the parameters $\theta_{CM}$ and $\theta_{CU}$ is
\begin{equation*}
    \mathcal{L}(\mathbf{C}^{st}|B^{st}=1,\mathbf{X}^{s},\mathbf{X}^{t},\theta_{CM},\theta_{CU})= \prod_{i^s =1}^{n_{1s}} \prod_{j^t =1}^{n_{2t}} \big[ f(\mathbf{\Gamma}_{Cij}^{st}|\theta_{CM})\big]^{C_{ij}^{st}} \big[f(\mathbf{\Gamma}_{Cij}^{st}|\theta_{CU})\big]^{1-C_{ij}^{st}}
\end{equation*}

Following the sampling algorithm proposed in \citet{Sadinle2017}, updates to the linkage structure are obtained by iterating through each entry $i^s \in s, s \in \mathbf{F}_1$ and proposing new link designations for each record $i^s$. At each iteration, there are two options: record $i^s$ forms a link with an unlinked record from block $t \in \mathbf{F}_2$, or record $i^s$ is not linked to any record. This algorithm preserves the one-to-one nature of $\mathbf{C}^{st}$ at every iteration. Let $\mathbf{C}_{-i}^{st}=(\mathbf{C}_{1,*}^{st}, \dots, \mathbf{C}_{i-1,*}^{st},\mathbf{C}_{i+1, *}^{st},\dots,\mathbf{C}_{n_{1s},*}^{st})^T$ be the linkage structure for $(s,t)$ excluding the designations belonging to $i^s$, and let $n_{m_{-i}}^{st}=\sum_{i^s=1}^{n_{1s}} \sum_{j^t=1}^{n_{2t}} \mathbf{C}_{{-i}}^{st}$ represent the number links in block $(s,t)$ excluding the link status of record $i^s$ from $s \in \mathbf{F}_1$. The posterior linking distribution when $i^s$ forms a true link with record $j\in t, t \in \mathbf{F}_2$ takes the form
\begin{align*}
    &f(C_{ij}^{st}=1, \mathbf{C}_{-i}^{st},n_{m}^{st}=n_{m_{-i}}^{st}+1| B^{st}=1,\mathbf{X}_{C}^s,\mathbf{X}_{C}^t,\theta_{CM},\theta_{CU}) \propto\\
 &p(\mathbf{C}^{st},n_m^{st}=n_{m_{-i}}^{st}+1|B^{st}=1) f(\mathbf{\Gamma}_{C ij}^{st}|\theta_{CM}) \mathbbm{1}\bigg(\sum_{i^s=1}^{n_{1s}} \mathbf{C}_{*,j}^{st}=0\bigg) \prod_{j'^t\neq j^t} f(\mathbf{\Gamma}_{C ij'}^{st}|\theta_{CU})
\end{align*}
where $\mathbbm{1}(\sum_{i^s}^{n_{1s}} \mathbf{C}_{*,j}^{st}=0)$ is an indicator for record $j \in t, t \in \mathbf{F}_2$ not being linked with any record in $s$. The posterior distribution when $i^s$ does not have a true link with any record in $t \in \mathbf{F}_2$ is
\begin{align*}
    &f(C_{i,*}^{st}=0, \mathbf{C}_{-i}^{st},n_{m}^{st}=n_{m_{-i}}^{st}|B^{st}=1,\mathbf{X}_{C}^s, \mathbf{X}_{C}^t,\theta_{CM},\theta_{CU}) \propto \\
     & p(\mathbf{C}^{st},n_m^{st}=n_{m_{-i}}^{st}|B^{st}=1) \prod_{j^t=1}^{n_{2t}}f(\mathbf{\Gamma}_{C i j}^{st}|\theta_{CU}).
\end{align*}
\normalsize
Updating the true link designation for $i^s$ is equivalent to sampling from a multinomial distribution of the possible link designations given the designations for the remaining records in $s \in \mathbf{F}_1$ do not change. Without a loss of generality, assume that the record size $n_{1s} < n_{2t}$ in block $(s,t)$. It can be shown that the probability for $i^s$ to pair with any record $j^t$ that does not have a true link is
\small{
\begin{align*}
    &P(C_{i j}^{st}=1|B^{st}=1,\mathbf{C}_{-i}^{st},\mathbf{X}_{C}^s,\mathbf{X}_{C}^t,\theta_{CM},\theta_{CU})= \\
    &\dfrac{\dfrac{P(\mathbf{\Gamma}_{C i\prime j\prime}^{st}|\theta_{CM})}{P(\mathbf{\Gamma}_{C i\prime j\prime}^{st}|\theta_{CU})}\mathbbm{1}(\sum_{i^{s}=1}^{n_{1s}} \mathbf{C}_{*,j}^{st}=0)}{\sum_{j^{t\prime}=1}^{n_{2t}}\dfrac{P(\mathbf{\Gamma}_{C i\prime j\prime}^{st}|\theta_{CM})}{P(\mathbf{\Gamma}_{C i\prime j\prime}^{st}|\theta_{CU})}\mathbbm{1}(\sum_{i^s=1}^{n_{1s}}\mathbf{C}_{*,j}^{st}=0)+\dfrac{(n_{2t}-n_{m_{-i}}^{st})(n_{1s}-n_{m_{-i}}^{st}+\beta_{\pi}-1)}{n_{m_{-i}}^{st}+\alpha_{\pi}}}
\end{align*}
}
\normalsize
The probability for $i_s$ to not pair with any record in $t \in \mathbf{F}_2$ is then
\small{
\begin{align*}
        &P(C_{i,*}^{st}=0|B^{st}=1,\mathbf{C}_{-i}^{st},\mathbf{X}_{C}^s,\mathbf{X}_{C}^t,\theta_{CM},\theta_{CU})= \\
    &\dfrac{\dfrac{(n_{2t}-n_{m_{-i}}^{st})(n_{1s}-n_{m_{-i}}^{st}+\beta_{\pi}-1)}{n_{m_{-i}}^{st}+\alpha_{\pi}}}{\sum_{j^{t\prime}=1}^{n_{2t}}\dfrac{P(\mathbf{\Gamma}_{C i' j'}^{st}|\theta_{CM})}{P(\mathbf{\Gamma}_{C i' j'}^{st}|\theta_{CU})}\mathbbm{1}(\sum_{i^s=1}^{n_{1s}}\mathbf{C}_{*, j}^{st}=0)+\dfrac{(n_{2t}-n_{m_{-i}}^{st})(n_{1s}-n_{m_{-i}}^{st}+\beta_{\pi}-1)}{n_{m_{-i}}^{st}+\alpha_{\pi}}}.
\end{align*}
}
\normalsize

\section{Appendix B: Metropolis Hastings Algorithm for Block Updates}
We describe the Metropolis Hastings updates to the blocking configuration $\mathbf{B}$. At iteration $[\nu]$, let block pair $(s,t), s \in \mathbf{F}_1, t \in \mathbf{F}_2$ represent a true block pair, let $\mathbf{C}^{st}$ denote the corresponding linkage structure within block pair $(s,t)$, and let $n_{m}^{st}$ be the number of estimated links within $(s,t)$. We can propose a new configuration $\mathbf{B}^{*}$ by sampling with equal probability block $r \in \mathbf{F}_2, r \neq t$ to link with $s \in \mathbf{F}_1$. Block $r$ can either be linked with block $q \in \mathbf{F}_1$, or not be linked to any blocks in $\mathbf{F}_1$. Depending on the link status of block r, two possible updates to $\mathbf{B}^{[v]}$ can be proposed.

The first possible update occurs when $r$ is not paired with any block in $\mathbf{F}_1$ at iteration $[v]$. Because we wish to preserve the complete one-to-one structure of the blocking configuration, this update will also involve a shift of block pair $(s,t)$ from $\mathbf{B}_M$ to $\mathbf{B}_U$ to ensure $s \in \mathbf{F}_1$ is only linked to one block in $\mathbf{F}_2$. At iteration $[v]$, block pair $(s,t)$ has an estimated linking configuration $\mathbf{C}^{st}$. In addition, all record pairs in block pair $(s,r)$ belong to the mixture $\mathbf{C}_{NB}$. The proposed update that shifts block pair $(s,r)$ into $\mathbf{B}_M$ and block pair $(s,t)$ into $\mathbf{B}_U$ assigns all record pairs from $(s,t)$ into $\mathbf{C}_{NB}$. In addition, record pairs in $(s,r)$ are assigned to either $\mathbf{C}_M$ or $\mathbf{C}_U$. Thus, it is necessary to propose a realization of $\mathbf{C}^{sr}|B^{sr}=1$ when updating the blocking status of $(s,r)$.

The second possible update occurs when $r \in \mathbf{F}_2$ is linked with $q \in \mathbf{F}_1$ at iteration $[v]$. We propose a swap in block linkage for $s,q \in \mathbf{F}_1$ such that we consider the new block pairs to be $(s,r)$ and $(q,t)$ at iteration $[v+1]$. At iteration $[v]$, the true block pairs $(s,t)$ and $(q,r)$ have linkage structures $\mathbf{C}^{st}$ and $\mathbf{C}^{qr}$, with $n_m^{st}$ and $n_m^{qr}$ records in $\mathbf{C}_M$, respectively. All record pairs associated with block pairs $(s,r)$ and $(q,t)$ are in $\mathbf{C}_{NB}$ at iteration $[\nu]$. The proposed update will assign $(s,r)$ and $(q,t)$ to $\mathbf{B}_M$ while assigning $(s,t)$ and $(q,r)$ to $\mathbf{B}_U$. The record pairs associated with $(s,r)$ and $(q,t)$ can be assigned to $\mathbf{C}_M$ or $\mathbf{C}_U$, and the record pairs belonging to $(s,t)$ and $(q,r)$ are all assigned to $\mathbf{C}_{NB}$. The linking states $\mathbf{C}^{sr}|B^{sr}=1$ and $\mathbf{C}^{qt}|B^{qt}=1$ need to be proposed along with the updates to the blocking structure.

One computationally efficient and stable method of proposing realizations of the record-level linkage state for block updates is to pre-specify a potential record-level linkage structure for all possible block pairs prior to implementing the MCMC sampling algorithm described in Section 2.3. To derive this potential record-level linkage structure, we estimated a suboptimal linkage structure for each pair of blocks. The EM algorithm was used to estimate the parameters of the \citet{Fellegi1969} model using record pairs from all possible pairs of blocks $s \in \mathbf{F}_1$ and $t \in \mathbf{F}_2$ (\cite{Winkler1989}). A one-to-one linking configuration was then estimated for all block pairs using the linear-sum assignment procedure (\cite{Jaro1989}). Let $\mathbf{C}^{st[pl]}$ be the pre-specified record-level linkage proposal for block $(s,t)$ and let $\mathbf{C}^{[pl]}=\{\mathbf{C}^{st[pl]}\}$ represent the collection of linkage proposals for $s=1, \dots, S$ and $t=1, \dots, T$. After block $(s,t)$ is sampled to the set $\mathbf{B}_M$ using the MCMC algorithm described in Section 2.3, values of $\mathbf{C}^{st[pl]}$ can be replaced with the sampled values of $\mathbf{C}^{st[\nu]}$ at every iteration after a burn-in period. This will adaptively update the set of proposed record-level linkage structures to be the most recently estimated linking configuration within each block pair.

Let $\mathbf{B}^{*}$ be a realization of the new block linking status and let $\mathbf{C}^{*}$ be its corresponding record-level linkage state. The Metropolis-Hastings acceptance probability for these block updates given the linkage structure is
\begin{equation*}
    A=\min \bigg(1, \dfrac{f(\mathbf{B}^{*},\mathbf{C}^{*}|\mathbf{X}_1,\mathbf{X}_2,\Theta) J(\mathbf{B},\mathbf{C}|\mathbf{B}^{*},\mathbf{C}^{*})}{f(\mathbf{B},\mathbf{C}|\mathbf{X}_1,\mathbf{X}_2,\Theta) J(\mathbf{B}^{*},\mathbf{C}^{*}|\mathbf{B},\mathbf{C})} \bigg)
\end{equation*}
where $f(\mathbf{B}^{*},\mathbf{C}^{*}|\mathbf{X}_1,\mathbf{X}_2,\Theta)$ represents the distribution for the proposed blocking and linking configuration, $f(\mathbf{B},\mathbf{C}|\mathbf{X}_1,\mathbf{X}_2,\Theta)$ represents the distribution for the current model state, $J(\mathbf{B}^{*},\mathbf{C}^{*}|\mathbf{B},\mathbf{C})$ is the transition probability to move to the new model state from the current model state, and $J(\mathbf{B},\mathbf{C}|\mathbf{B}^{*},\mathbf{C}^{*})$ is the transition probability for the reverse move. It is useful to express the joint posterior distribution for $\mathbf{B}$ and $\mathbf{C}$ as 
\begin{equation*}
    f(\mathbf{B},\mathbf{C}|\mathbf{X}_1,\mathbf{X}_2,\Theta) \propto p(\mathbf{B}) p(\mathbf{C}|\mathbf{B}) \mathcal{L} (\mathbf{B},\mathbf{C}|\mathbf{X}_1, \mathbf{X}_2,\Theta).
\end{equation*}
The following sections provide more details about the form of the acceptance probability for the two types of block pair updates using proposals for $\mathbf{C}^{*}|\mathbf{B}^{*}$ from $\mathbf{C}^{[pl]}$.

\begin{comment}
Similarly, the transition probability for the proposed (and reverse) model state can be factored into a blocking and a linking component
\begin{equation*}
    J(\mathbf{B}^{*},\mathbf{C}^{*}|\mathbf{B},\mathbf{C})=J(\mathbf{B}^{*}|\mathbf{B},\mathbf{C}) J(\mathbf{C}^{*}|\mathbf{B}^{*},\mathbf{B},\mathbf{C}).
\end{equation*}
where $J(\mathbf{B}^{*}|\mathbf{B},\mathbf{C})$ represents the transition probability for choosing the update to the blocking configuration and $J(\mathbf{C}^{*}|\mathbf{B}^{*},\mathbf{B},\mathbf{C})$ represents the probability for selecting the proposed linking configuration given the proposed blocking state. 
\end{comment}

\subsection{Move Type 1}
We first consider the scenario where $r \in \mathbf{F}_2$ at iteration $[v]$ is not linked to any block in $\mathbf{F}_1$. We propose an update to $\mathbf{B}$ that assigns block pair $(s,r)$ into $\mathbf{B}_M$ while assigning $(s,t)$ into $\mathbf{B}_U$. All record pairs in $(s,t)$ are categorized as non-links, while $\mathbf{C}^{sr[pl]}$ is proposed as the linkage estimate for $(s,r)$. The blocking configuration at iteration $[v]$ can be expressed as $ B^{st[\nu]}=1, B^{sr[\nu]}=0$ and the linking configuration can be written as $\mathbf{C}^{st[\nu]}; \mathbf{C}^{sr[\nu]}=0, \forall i^s=1, \dots, n_{1s}, j^r =1, \dots, n_{2r}$. The proposed update can be expressed as $ B^{st[*]}=0, B^{sr[*]}=1$ and $\mathbf{C}^{st[*]}=0, \forall i_s= 1, \dots, n_{1s}, j_t =1, \dots, n_{2t}; \mathbf{C}^{sr[*]}=\mathbf{C}^{sr[pl]}$.

The probability of selecting $\mathbf{B}^{*}$ as the update to the blocking configuration and $\mathbf{C}^{*}$ as the corresponding linking configuration is the probability of linking $r$ to $s$, which is uniform over the blocks in $\mathbf{F}_2$ that are not linked to any other block at $[\nu]$. This is equal to $J(\mathbf{B}^{*},\mathbf{C}^{*}|\mathbf{B},\mathbf{C})=1/(T-S)$. The transition probability for the reverse move is the probability of selecting block $t\in \mathbf{F}_2$ that is not linked to any blocks, which is also equal to $J(\mathbf{B},\mathbf{C}|\mathbf{B}^{*},\mathbf{C}^{*})=1/(T-S)$. 

\begin{comment}
The record-level linking component of the transition probability is the probability of obtaining the proposed linking configuration $\mathbf{C}^{sr}$ with $n_{m}^{st}$ true link record pairs, given $(s,r)$ is a true block pair. This probability is equal to
\begin{align*}
    J(\mathbf{C}^{[v+1]}|\mathbf{B}^{[v+1]},\mathbf{B}^{[v]},\mathbf{C}^{[v]})=&p(n_{m}^{sr[v+1]}|\pi)p(\mathbf{C}^{sr[v+1]}|n_{m}^{sr[v+1]})\\
    = & \binom{\min(n_{1s},n_{2r})}{n_{m}^{sr[v+1]}} \pi^{n_{m}^{sr[v+1]}}(1-\pi)^{\min(n_{1s},n_{2r})-n_{m}^{sr[v+1]}} \times \binom{n_{1s}}{n_{m}^{sr[v+1]}} \binom{n_{2r}}{n_{m}^{sr[v+1]}} n_{m}^{sr[v+1]}!.
\end{align*}
All entries in $\mathbf{C}^{st}$ are set to 0 with probability 1 in the proposed update. The transition probability of the linkage structure for the reverse move is the probability of obtaining the original linkage structure $\mathbf{C}^{st}$ with $n_m^{st}$ true links. This is equal to
\begin{align*}
    J(\mathbf{C}^{[v]}|\mathbf{B}^{[v]},\mathbf{B}^{[v+1]},\mathbf{C}^{[v+1]})=& p(n_{m}^{st[v]}|\pi)p(\mathbf{C}^{st[v]}|n_{m}^{st[v]})\\
    = &\binom{\min(n_{1s},n_{2t})}{n_{m}^{st[v]}} \pi^{n_{m}^{st[v]}}(1-\pi)^{\min(n_{1s},n_{2t})-n_{m}^{st[v]}} \times \binom{n_{1s}}{n_{m}^{st[v]}} \binom{n_{2t}}{n_{m}^{st[v]}} n_{m}^{st[v]}!
\end{align*}
with all entries in $\mathbf{C}^{sr}$ equal to 0 with probability 1.
\end{comment}

The prior distributions for both the proposed and original blocking states are uniform over all possible blocking configurations that are complete and one-to-one, which is equal to $p(\mathbf{B}^{*})=p(\mathbf{B})=\binom{S}{S} \binom{T}{S} S!$. The prior distribution for the proposed linking state $\mathbf{C}^{*}|\mathbf{B}^{*}$ is the product of independent prior distributions for the linking states within all block pairs according to $\mathbf{B}^{*}$ with $B^{sr[*]}=1$. The prior distribution of the linkage structure within true block pairs takes the form of Equation \eqref{8}, while the prior distribution of the linkage structure for non-block pairs is a point mass at 0 with probability 1. This can be expressed as 
\begin{align*}
    p(\mathbf{C}^{*}|\mathbf{B}^{*})= &\prod_{s=1}^S \prod_{t=1}^T p(\mathbf{C}^{st[*]},n_m^{st[*]}|\alpha_{\pi}, \beta_{\pi})^{\mathbbm{1}(B^{st[*]}=1)} p(\mathbf{C}^{st[*]}=0)^{\mathbbm{1}(B^{st[*]}=0)}\\
    =& p(\mathbf{C}^{sr[pl]},n_m^{sr[pl]}|\alpha_{\pi},\beta_{\pi}) \prod_{s't' \neq sr} p(\mathbf{C}^{s't'[\nu]},n_m^{s't'[\nu]}|\alpha_{\pi},\beta_{\pi})^{\mathbbm{1}(B^{s't'[*]}=1)}.
\end{align*}
Similarly, the prior distribution for the original linking state is the product of independent linking priors for block pairs according to $\mathbf{B}^{[\nu]}$ with $B^{st[\nu]}=1$ is
\begin{align*}
        p(\mathbf{C}|\mathbf{B})= &\prod_{s=1}^S \prod_{t=1}^T p(\mathbf{C}^{st[\nu]},n_m^{st[\nu]}|\alpha_{\pi}, \beta_{\pi})^{\mathbbm{1}(B^{st[\nu]}=1)} p(\mathbf{C}^{st[\nu]}=0)^{\mathbbm{1}(B^{st[\nu]}=0)}\\
    =& p(\mathbf{C}^{st[\nu]},n_m^{st[\nu]}|\alpha_{\pi},\beta_{\pi}) \prod_{s't' \neq st} p(\mathbf{C}^{s't'[\nu]},n_m^{s't'[\nu]}|\alpha_{\pi},\beta_{\pi})^{\mathbbm{1}(B^{s't'[\nu]}=1)}.
\end{align*}
The ratio of prior distributions $p(\mathbf{C}^{*}|\mathbf{B}^{*})/p(\mathbf{C}|\mathbf{B})$ reduces to \\ $p(\mathbf{C}^{sr[pl]},n_m^{sr[pl]}|\alpha_{\pi},\beta_{\pi}) / p(\mathbf{C}^{st[\nu]},n_m^{st[\nu]}|\alpha_{\pi},\beta_{\pi})$, which is equal to
\begin{align*}
&\dfrac{p(\mathbf{C}^{sr[pl]},n_m^{sr[pl]}|\alpha_{\pi},\beta_{\pi})}{p(\mathbf{C}^{st[\nu]},n_m^{st[\nu]}|\alpha_{\pi},\beta_{\pi})}\\
=& \dfrac{\dfrac{(\max(n_{1s},n_{2r})-n_m^{sr[pl]})!}{\max(n_{1s},n_{2r})!}\times \dfrac{\Gamma(n_m^{sr[pl]}+\alpha_{\pi})\Gamma(\min(n_{1s},n_{2r})-n_m^{sr[pl]}+\beta_{\pi})}{\Gamma(\min(n_{1s},n_{2r})+\alpha_{\pi}+\beta_{\pi})}}{\dfrac{(\max(n_{1s},n_{2t})-n_m^{st[\nu]})!}{\max(n_{1s},n_{2t})!}\times \dfrac{\Gamma(n_m^{st[\nu]}+\alpha_{\pi})\Gamma(\min(n_{1s},n_{2t})-n_m^{st[\nu]}+\beta_{\pi})}{\Gamma(\min(n_{1s},n_{2t})+\alpha_{\pi}+\beta_{\pi})}}
\end{align*}

The ratio of joint likelihoods for the proposed and original blocking and linking states is
\begin{align*}
    &\dfrac{\mathcal{L}(\mathbf{B}^{*},\mathbf{C}^{*}| \mathbf{X}_1,\mathbf{X}_2, \Theta)}{\mathcal{L}(\mathbf{B},\mathbf{C}|\mathbf{X}_1,\mathbf{X}_2, \Theta) }\\
    = & \dfrac{\mathcal{L}(B^{sr[*]}=1,  B^{st[*]}=0, \mathbf{C}^{sr[pl]}, \mathbf{C}^{st[*]}=0|\mathbf{X}^{s}, \mathbf{X}^{t}, \mathbf{X}^{r},\Theta)\prod_{s't' \neq \{st,sr \}} \mathcal{L}(B^{s't'[\nu]}, \mathbf{C}^{s' t'[\nu]}|\mathbf{X}^{s'},\mathbf{X}^{t'},\Theta)}{\mathcal{L}(B^{st[\nu]}=1,  B^{sr[\nu]}=0,\mathbf{C}^{st[\nu]},\mathbf{C}^{sr[\nu]}=0|\mathbf{X}^{s},\mathbf{X}^{t},\mathbf{X}^{r},\Theta)\prod_{s't' \neq \{st,sr \}} \mathcal{L}(B^{s't'[\nu]}, \mathbf{C}^{s' t'[\nu]}|\mathbf{X}^{s'},\mathbf{X}^{t'},\Theta)}\\
    =& \dfrac{f(\mathbf{\Gamma}_{B}^{sr}|\theta_{BM})f(\mathbf{\Gamma}_{C}^{sr}|\theta_{CM},\theta_{CU})f(\mathbf{\Gamma}_{B}^{st}|\theta_{BU})f(\mathbf{\Gamma}_{C}^{st}|\theta_{CNB})}{f(\mathbf{\Gamma}_{B}^{st}|\theta_{BM})f(\mathbf{\Gamma}_{C}^{st}|\theta_{CM},\theta_{CU})f(\mathbf{\Gamma}_{B}^{sr}|\theta_{BU})f(\mathbf{\Gamma}_{C}^{sr}|\theta_{CNB})}
\end{align*}
Only the blocking and linking information for block pairs $(s,t)$ and $(s,r)$ appear in this ratio because the information in the other block pairs remain the same for the proposed and original states. Under the likelihood in Equation \eqref{6}, the ratio of joint likelihoods can be written explicitly as
\begin{align*}
&\dfrac{\mathcal{L}(B^{st[*]},B^{sr[*]},\mathbf{C}^{st[*]},\mathbf{C}^{sr[pl]}|\mathbf{X}^s,\mathbf{X}^r,\mathbf{X}^t, \Theta)}{\mathcal{L}(B^{st[\nu]},B^{sr[\nu]},\mathbf{C}^{st[\nu]},\mathbf{C}^{sr[\nu]}|\mathbf{X}^s,\mathbf{X}^r,\mathbf{X}^t, \Theta) }\\
=&\dfrac{\prod_{p=1}^P \theta_{BMp}^{\Gamma_{Bp}^{sr}}(1-\theta_{BMp})^{1-\Gamma_{Bp}^{sr}}}{\prod_{p=1}^P \theta_{BMp}^{\Gamma_{Bp}^{st}}(1-\theta_{BMp})^{1-\Gamma_{Bp}^{st}}}\\
    \times & \dfrac{\prod_{i^s=1}^{n_{1s}} \prod_{j^r=1}^{n_{2r}} \prod_{k=1}^K\{ \theta_{CMk}^{\Gamma_{Cij k}^{sr}}(1-\theta_{CMk})^{1-\Gamma_{Cijk}^{sr}} \}^{C_{ij}^{sr}}\{ \theta_{CUk}^{\Gamma_{Cijk}^{sr}}(1-\theta_{CUk})^{1-\Gamma_{Cijk}^{sr}}\}^{1-C_{ij}^{sr}}}{\prod_{i^s=1}^{n_{1s}} \prod_{j^t=1}^{n_{2t}} \prod_{k=1}^K\{ \theta_{CMk}^{\Gamma_{Cijk}^{st}}(1-\theta_{CMk})^{1-\Gamma_{Cijk}^{st}} \}^{C_{ij}^{st}}\{ \theta_{CUk}^{\Gamma_{Cijk}^{st}}(1-\theta_{CUk})^{1-\Gamma_{Cijk}^{st}}\}^{1-C_{ij}^{st}}}\\
    \times & \dfrac{\prod_{p=1}^P \theta_{BUp}^{\Gamma_{Bp}^{st}}(1-\theta_{BUp})^{1-\Gamma_{Bp}^{st}}}{\prod_{p=1}^P \theta_{BUp}^{\Gamma_{Bp}^{sr}}(1-\theta_{BUp})^{1-\Gamma_{Bp}^{sr}}}\times \dfrac{\prod_{i^s=1}^{n_{1s}} \prod_{j^t=1}^{n_{2t}} \prod_{k=1}^K\{ \theta_{CNBk}^{\Gamma_{Cijk}^{st}}(1-\theta_{CNBk})^{1-\Gamma_{Cijk}^{st}} \}}{\prod_{i^s=1}^{n_{1s}} \prod_{j^r=1}^{n_{2r}} \prod_{k=1}^K\{ \theta_{CNBk}^{\Gamma_{Cijk}^{sr}}(1-\theta_{CNBk})^{1-\Gamma_{Cijk}^{sr}} \}}
\end{align*}

Therefore, the acceptance probability for this Metropolis-Hastings update reduces to the ratio of prior linking distributions and the joint likelihoods for the proposed and original blocking and linking states of $(s,t)$ and $(s,r)$:
\begin{equation*}
    A=\min \bigg(1, \dfrac{p(\mathbf{C}^{sr[pl]},n_m^{sr[pl]}|\alpha_{\pi},\beta_{\pi})}{p(\mathbf{C}^{st[\nu]},n_m^{st[\nu]}|\alpha_{\pi},\beta_{\pi})} \times \dfrac{\mathcal{L}(B^{st[*]},B^{sr[*]},\mathbf{C}^{st[*]},\mathbf{C}^{sr[pl]}|\mathbf{X}^s,\mathbf{X}^r,\mathbf{X}^t, \Theta)}{\mathcal{L}(B^{st[\nu]},B^{sr[\nu]},\mathbf{C}^{st[\nu]},\mathbf{C}^{sr[\nu]}|\mathbf{X}^s,\mathbf{X}^r,\mathbf{X}^t, \Theta) } \bigg)
\end{equation*}

\subsection{Move Type 2}
This move is applied when block $r \in \mathbf{F}_2$ is linked with $q \in \mathbf{F}_1$ at iteration $[\nu]$. The blocking configuration at iteration $[\nu]$ can be written as $B^{st[\nu]}=1, B^{qr[\nu]}=1, B^{sr[\nu]}=0, B^{qt[\nu]}=0$ and the linking configuration is $\mathbf{C}^{st[\nu]}; \mathbf{C}^{qr[\nu]}; \mathbf{C}^{sr[\nu]}=0, \forall i^s=1, \dots, n_{1s}, j^r=1, \dots, n_{2r}; \mathbf{C}^{qt[\nu]}=0, \forall i^q=1, \dots, n_{1q}, j^t=1, \dots, n_{2t} $. The proposed block update can be expressed as $B^{st[*]}=0, B^{qr[*]}=0, B^{sr[*]}=1, B^{qt[*]}=1$ and the linking configuration update is $\mathbf{C}^{st[*]}=0, \forall i_s=1, \dots, n_{1s}; \mathbf{C}^{qr[*]}=0, \forall i_q=1, \dots, n_{1q}, j_r=1, \dots, n_{2r};\mathbf{C}^{sr[*]}=\mathbf{C}^{sr[pl]}; \mathbf{C}^{qt[*]}=\mathbf{C}^{qt[pl]}$.

The transition probability of selecting $\mathbf{B}^{*}$ as the update to the blocking configuration and $\mathbf{C}^{*}$ as the linking configuration is the probability of selecting the block pair $(q,r) \in \mathbf{B_M}$ to swap designations with $(s,t)$, which is equal to $J(\mathbf{B}^{*},\mathbf{C}^{*}|\mathbf{B},\mathbf{C})=1/(S-1)$. The transition probability for the reverse move is the probability of selecting the block pair $(q,t) \in \mathbf{B_M}$ to swap designations with $(s,r)$, which is also equal to $p(\mathbf{B},\mathbf{C}|\mathbf{B}^{*},\mathbf{C}^{*})=1/(S-1)$.

As with the first type of updates, the prior distributions for the proposed and original blocking states are uniform over all possible blocking configurations that are complete and one-to-one, which is expressed as $p(\mathbf{B}^{*})=p(\mathbf{B})=\binom{S}{S} \binom{T}{S} S!$. The prior distribution for the proposed linking state $\mathbf{C}^{*}|\mathbf{B}^{*}$ is the product of independent prior distributions for the linking states within all block pairs according to $\mathbf{B}^{*}$, which proposes $B^{sr[*]}=1$ and $B^{qt[*]}=1$. The prior distribution for linking states among block pairs in $\mathbf{B_M}$ take the form of Equation \eqref{8}, while prior distributions among blocks in $\mathbf{B_U}$ have a point mass at 0 with probability 1. This can be expressed as
\begin{align*}
    p(\mathbf{C}^{*}|\mathbf{B}^{*})= &\prod_{s=1}^S \prod_{t=1}^T p(\mathbf{C}^{st[*]},n_m^{st[*]}|\alpha_{\pi}, \beta_{\pi})^{\mathbbm{1}(B^{st[*]}=1)} p(\mathbf{C}^{st[*]}=0)^{\mathbbm{1}(B^{st[*]}=0)}\\
    =& p(\mathbf{C}^{sr[pl]},n_m^{sr[pl]}|\alpha_{\pi},\beta_{\pi}) p(\mathbf{C}^{qt[pl]},n_m^{qt[pl]}|\alpha_{\pi},\beta_{\pi}) \\
    \times &\prod_{s't' \neq \{sr,qt \}} p(\mathbf{C}^{s't'[\nu]},n_m^{s't'[\nu]}|\alpha_{\pi},\beta_{\pi})^{\mathbbm{1}(B^{s't'[*]}=1)}.
\end{align*}
Similarly, the prior distribution for the original linking state is the product of independent prior distributions for linkage structures of block pairs according to $\mathbf{B}^{[\nu]}$, which retains $B^{st[\nu]}=1$ and $B^{qr[\nu]}=1$. This is equal to
\begin{align*}
        p(\mathbf{C}|\mathbf{B})= &\prod_{s=1}^S \prod_{t=1}^T p(\mathbf{C}^{st[\nu]},n_m^{st[\nu]}|\alpha_{\pi}, \beta_{\pi})^{\mathbbm{1}(B^{st[\nu]}=1)} p(\mathbf{C}^{st[\nu]}=0)^{\mathbbm{1}(B^{st[\nu]}=0)}\\
    =& p(\mathbf{C}^{st[\nu]},n_m^{st[\nu]}|\alpha_{\pi},\beta_{\pi}) p(\mathbf{C}^{qr[\nu]},n_m^{qr[\nu]}|\alpha_{\pi},\beta_{\pi})  \\
    \times &\prod_{s't' \neq \{st, qr \}} p(\mathbf{C}^{s't'[\nu]},n_m^{s't'[\nu]}|\alpha_{\pi},\beta_{\pi})^{\mathbbm{1}(B^{s't'[\nu]}=1)}.
\end{align*}
The ratio of prior distributions $p(\mathbf{C}^{*}|\mathbf{B}^{*})/p(\mathbf{C}|\mathbf{B})$ reduces to 
\begin{align*}
&\dfrac{p(\mathbf{C}^{sr[pl]},n_m^{sr[pl]}|\alpha_{\pi},\beta_{\pi})p(\mathbf{C}^{qt[pl]},n_m^{qt[pl]}|\alpha_{\pi},\beta_{\pi})}{p(\mathbf{C}^{st[\nu]},n_m^{st[\nu]}|\alpha_{\pi},\beta_{\pi})p(\mathbf{C}^{qr[\nu]},n_m^{qr[\nu]}|\alpha_{\pi},\beta_{\pi})}\\
=& \dfrac{\dfrac{(\max(n_{1s},n_{2r})-n_m^{sr[pl]})!}{\max(n_{1s},n_{2r})!}\times \dfrac{\Gamma(n_m^{sr[pl]}+\alpha_{\pi})\Gamma(\min(n_{1s},n_{2r})-n_m^{sr[pl]}+\beta_{\pi})}{\Gamma(\min(n_{1s},n_{2r})+\alpha_{\pi}+\beta_{\pi})}}{\dfrac{(\max(n_{1s},n_{2t})-n_m^{st[\nu]})!}{\max(n_{1s},n_{2t})!}\times \dfrac{\Gamma(n_m^{st[\nu]}+\alpha_{\pi})\Gamma(\min(n_{1s},n_{2t})-n_m^{st[\nu]}+\beta_{\pi})}{\Gamma(\min(n_{1s},n_{2t})+\alpha_{\pi}+\beta_{\pi})}} \\
\times & \dfrac{\dfrac{(\max(n_{1q},n_{2t})-n_m^{qt[pl]})!}{\max(n_{1q},n_{2t})!}\times \dfrac{\Gamma(n_m^{qt[pl]}+\alpha_{\pi})\Gamma(\min(n_{1q},n_{2t})-n_m^{qt[pl]}+\beta_{\pi})}{\Gamma(\min(n_{1q},n_{2t})+\alpha_{\pi}+\beta_{\pi})}}{\dfrac{(\max(n_{1q},n_{2r})-n_m^{qr[\nu]})!}{\max(n_{1q},n_{2r})!}\times \dfrac{\Gamma(n_m^{qr[\nu]}+\alpha_{\pi})\Gamma(\min(n_{1q},n_{2r})-n_m^{qr[\nu]}+\beta_{\pi})}{\Gamma(\min(n_{1q},n_{2r})+\alpha_{\pi}+\beta_{\pi})}}.
\end{align*}

The ratio of joint likelihoods for the proposed and original blocking and linking states is
This ratio is
\begin{align*}
    &\dfrac{\mathcal{L}(\mathbf{B}^{*},\mathbf{C}^{*}|\mathbf{X}_1,\mathbf{X}_2,\Theta)}{\mathcal{L}(\mathbf{B},\mathbf{C}|\mathbf{X}_1,\mathbf{X}_2,\Theta)} \\
    = & \dfrac{\mathcal{L}(B^{sr[*]}=1,B^{qt[*]}=1,B^{st[*]}=0,B^{qr[*]}=0,\mathbf{C}^{sr[pl]},\mathbf{C}^{qt[pl]},\mathbf{C}^{st[*]}=0,\mathbf{C}^{qr[*]}=0|\mathbf{X}^{s},\mathbf{X}^{q},\mathbf{X}^{t},\mathbf{X}^{r},\Theta)}{\mathcal{L}(B^{st[\nu]}=1,B^{qr[\nu]}=1,B^{sr[\nu]}=0,B^{qt[\nu]}=0,\mathbf{C}^{st[\nu]},\mathbf{C}^{qr[\nu]},\mathbf{C}^{sr[\nu]}=0,\mathbf{C}^{qt[\nu]}=0|\mathbf{X}^{s},\mathbf{X}^{q},\mathbf{X}^{t},\mathbf{X}^{r},\Theta)}  \\
    \times & \dfrac{\prod_{s't' \neq \{st, sr, qr, qt \}} \mathcal{L}(B^{s't'},\mathbf{C}^{s't'}|\mathbf{X}^{s'},\mathbf{X}^{t'},\Theta)}{\prod_{s't' \neq \{st, sr, qr, qt \}} \mathcal{L}(B^{s't'},\mathbf{C}^{s't'}|\mathbf{X}^{s'},\mathbf{X}^{t'},\Theta)}\\
    = & \dfrac{f(\mathbf{\Gamma}_{B}^{sr}|\theta_{BM}) f(\mathbf{\Gamma}_{C}^{sr}|\theta_{CM},\theta_{CU}) f(\mathbf{\Gamma}_{B}^{qt}|\theta_{BM}) f(\mathbf{\Gamma}_{C}^{qt}|\theta_{CM},\theta_{CU})}{f(\mathbf{\Gamma}_{B}^{st}|\theta_{BM}) f(\mathbf{\Gamma}_{C}^{st}|\theta_{CM},\theta_{CU}) f(\mathbf{\Gamma}_{B}^{qr}|\theta_{BM})f(\mathbf{\Gamma}_{C}^{qr}|\theta_{CM},\theta_{CU})} \\
    \times & \dfrac{f(\mathbf{\Gamma}_{B}^{st}|\theta_{BU}) f(\mathbf{\Gamma}_{C}^{st}|\theta_{CNB}) f(\mathbf{\Gamma}_{B}^{qr}|\theta_{BU}) f(\mathbf{\Gamma}_{C}^{qr}|\theta_{CNB})}{f(\mathbf{\Gamma}_{B}^{sr}|\theta_{BU}) f(\mathbf{\Gamma}_{C}^{sr}|\theta_{CNB}) f(\mathbf{\Gamma}_{C}^{qt}|\theta_{BU}) f(\mathbf{\Gamma}_{C}^{qt}|\theta_{CNB})}
\end{align*}
We see that the ratio of joint likelihoods for the proposed update and the original blocking and linking state reduces to a ratio of the blocking information for the true block pairs and non-block pairs, the linking likelihood among true block pairs, and the non-linking likelihood among non-block pairs. Using the form of the likelihood in (6), we can explicitly express the ratio of joint likelihoods as 

\begin{align*}
&\dfrac{\mathcal{L}(B^{st[*]},B^{sr[*]},B^{qt[*]},B^{qr[*]},\mathbf{C}^{st[*]},\mathbf{C}^{sr[pl]},\mathbf{C}^{qt[pl]},\mathbf{C}^{qr[*]}|\mathbf{X}^q,\mathbf{X}^s,\mathbf{X}^r,\mathbf{X}^t,\Theta)}{\mathcal{L}(B^{st[\nu]},B^{sr[\nu]},B^{qt[\nu]},B^{qr[\nu]},\mathbf{C}^{st[\nu]},\mathbf{C}^{sr[\nu]},\mathbf{C}^{qt[\nu]},\mathbf{C}^{qr[\nu]}|\mathbf{X}^q,\mathbf{X}^s,\mathbf{X}^r,\mathbf{X}^t\Theta)}\\
    = & \dfrac{\prod_{p=1}^P \theta_{BMp}^{\Gamma_{Bp}^{sr}}(1-\theta_{BMp})^{1-\Gamma_{Bp}^{sr}} \theta_{BMp}^{\Gamma_{Bp}^{qt}}(1-\theta_{BMp})^{1-\Gamma_{Bp}^{qt}}}{\prod_{p=1}^P \theta_{BMp}^{\Gamma_{Bp}^{st}}(1-\theta_{BMp})^{1-\Gamma_{Bp}^{st}} \theta_{BMp}^{\Gamma_{Bp}^{qr}}(1-\theta_{BMp})^{1-\Gamma_{Bp}^{qr}}}\\
    \times & \dfrac{\prod_{i^s=1}^{n_{1s}} \prod_{j^r=1}^{n_{2r}} \prod_{k=1}^K\{ \theta_{CMk}^{\Gamma_{Cijk}^{sr}}(1-\theta_{CMk})^{1-\Gamma_{Cijk}^{sr}} \}^{C_{ij}^{sr}}\{ \theta_{CUk}^{\Gamma_{Cijk}^{sr}}(1-\theta_{CUk})^{1-\Gamma_{Cijk}^{sr}}\}^{1-C_{ij}^{sr}}}{\prod_{i^s=1}^{n_{1s}} \prod_{j^t=1}^{n_{2t}} \prod_{k=1}^K\{ \theta_{CMk}^{\Gamma_{Cijk}^{st}}(1-\theta_{CMk})^{1-\Gamma_{Cijk}^{st}} \}^{C_{ij}^{st}}\{ \theta_{CUk}^{\Gamma_{Cijk}^{st}}(1-\theta_{CUk})^{1-\Gamma_{Cijk}^{st}}\}^{1-C_{ij}^{st}}}\\
    \times & \dfrac{\prod_{i^q=1}^{n_{1q}} \prod_{j^t=1}^{n_{2t}} \prod_{k=1}^K \{ \theta_{CMk}^{\Gamma_{Cijk}^{qt}}(1-\theta_{CMk})^{1-\Gamma_{Cijk}^{qt}} \}^{C_{ij}^{qt}}\{ \theta_{CUk}^{\Gamma_{Cijk}^{qt}}(1-\theta_{CUk})^{1-\Gamma_{C ijk}^{qt}} \}^{1-C_{ij}^{qt}}}{\prod_{i^q=1}^{n_{1q}} \prod_{j^r=1}^{n_{2r}}\prod_{k=1}^K \{ \theta_{CMk}^{\Gamma_{Ci jk}^{qr}}(1-\theta_{CMk})^{1-\Gamma_{Ci jk}^{qr}} \}^{C_{ij}^{qr}}\{ \theta_{CUk}^{\Gamma_{Cijk}^{qr}}(1-\theta_{CUk})^{1-\Gamma_{Ci jk}^{qr}} \}^{1-C_{ij}^{qr}}}\\
    \times & \dfrac{\prod_{p=1}^P \theta_{BUp}^{\Gamma_{Bp}^{st}}(1-\theta_{BUp})^{1-\Gamma_{Bp}^{st}} \theta_{BUp}^{\Gamma_{Bp}^{qr}} (1-\theta_{BUp})^{1-\Gamma_{Bp}^{qr}}}{\prod_{p=1}^P \theta_{BUp}^{\Gamma_{Bp}^{sr}}(1-\theta_{BUp})^{1-\Gamma_{Bp}^{sr}} \theta_{BUp}^{\Gamma_{Bp}^{qt}} (1-\theta_{BUp})^{1-\Gamma_{Bp}^{qt}}}\\
    \times & \dfrac{\prod_{i^s=1}^{n_{1s}} \prod_{j^t=1}^{n_{2t}} \prod_{k=1}^K\{ \theta_{CNBk}^{\Gamma_{Cijk}^{st}}(1-\theta_{CNBk})^{1-\Gamma_{Cijk}^{st}} \}}{\prod_{i^s=1}^{n_{1s}} \prod_{j^r=1}^{n_{2r}} \prod_{k=1}^K\{ \theta_{CNBk}^{\Gamma_{Cijk}^{sr}}(1-\theta_{CNBk})^{1-\Gamma_{Cijk}^{sr}} \}}
     \times  \dfrac{\prod_{i^q=1}^{n_{1q}} \prod_{j^r=1}^{n_{2r}}\prod_{k=1}^K  \{\theta_{CNBk}^{\Gamma_{Cijk}^{qr}}(1-\theta_{CNBk})^{1-\Gamma_{Cijk}^{qr}} \}}{\prod_{i^q=1}^{n_{1q}} \prod_{j^t=1}^{n_{2t}} \prod_{k=1}^K\{ \theta_{CNBk}^{\Gamma_{C ijk}^{qt}}(1-\theta_{CNBk})^{1-\Gamma_{C ijk}^{qt}} \}}.
\end{align*}

As a result, the Metropolis-Hastings acceptance probability for this proposed update simplifies to the ratio of prior linking distributions and the joint likelihoods for the proposed and original states of $(s,t), (s,r), (qr)$ and $(q,t)$:
\begin{align*}
    A=\min \bigg(1, &\dfrac{p(\mathbf{C}^{sr[pl]},n_m^{sr[pl]}|\alpha_{\pi},\beta_{\pi})p(\mathbf{C}^{qt[pl]},n_m^{qt[pl]}|\alpha_{\pi},\beta_{\pi})}{p(\mathbf{C}^{st[\nu]},n_m^{st[\nu]}|\alpha_{\pi},\beta_{\pi})p(\mathbf{C}^{qr[\nu]},n_m^{qr[\nu]}|\alpha_{\pi},\beta_{\pi})} \times\\ &\dfrac{\mathcal{L}(B^{st[*]},B^{sr[*]},B^{qt[*]},B^{qr[*]},\mathbf{C}^{st[*]},\mathbf{C}^{sr[pl]},\mathbf{C}^{qt[pl]},\mathbf{C}^{qr[*]}|\mathbf{X}^q,\mathbf{X}^s,\mathbf{X}^r,\mathbf{X}^t,\Theta)}{\mathcal{L}(B^{st[\nu]},B^{sr[\nu]},B^{qt[\nu]},B^{qr[\nu]},\mathbf{C}^{st[\nu]},\mathbf{C}^{sr[\nu]},\mathbf{C}^{qt[\nu]},\mathbf{C}^{qr[\nu]}|\mathbf{X}^q,\mathbf{X}^s,\mathbf{X}^r,\mathbf{X}^t\Theta)} \bigg).
\end{align*}

\section{Appendix C: Creation of Blocking and Linking Variables to link NTDB and MedPar datasets}
Trauma hospital linking characteristics include: hospital region, bed size, hospital trauma level, and the presence of pediatric beds. In addition, we derived block-level variables that indicate the presence of rare or severe traumatic injuries and procedures commonly associated with these injuries using the ICD-9-CM codes for all hospitalization events occurring at that facility among individuals 65 and older in 2015. These codes were identified in the NTDB and Medicare files using patient level information. The variables were defined as 1 if at least one patient record treated at the hospital included these codes, and 0 otherwise. Appendix Table 7 details the diagnosis and procedure codes that were derived from ICD-9-CM codes, as well as the codes used to identify traumatic brain injury.

\begin{table}[]
\caption*{Appendix Table 7: Description of ICD-9-CM Diagnosis and Procedure Variables}
\begin{tabular}{ll}
\hline
ICD-9 Diagnosis Code & Description \\ \hline
\begin{tabular}[c]{@{}l@{}l@{}} 310.0, 800.00-804.99,\\
850.00-854.99, 873.0, \\ 873.1, 905.0, 907.0, 959.01\\
\end{tabular} & Traumatic Brain Injury\\

801.7 & \begin{tabular}[c]{@{}l@{}}Open fracture of base of skull w/ subarachnoid, subdural, \\ and extradural hemorrhage\end{tabular} \\
801.5-801.9 & Open fracture of base of skull \\
802.1 & Open fracture of nasal bones \\
802.3 & Open fracture of mandible \\
802.5 & Open fracture of malar and maxillary bone \\
806.0 & Closed fracture of cervical vertebra w/ spinal cord injury \\
806.4 & Closed fracture of lumbar spine w/ spinal cord injury \\
812.1, 812.3, 812.5 & Open fracture of humerus \\
851.4 & \begin{tabular}[c]{@{}l@{}}Cerebellar or brain stem contusion w/o mention of open\\   intracranial wound\end{tabular} \\
880.0 & \begin{tabular}[c]{@{}l@{}}Open wound of shoulder and upper arm w/o mention of\\   complication\end{tabular} \\ \hline
ICD-9 Procedure Code & Description \\ \hline
1.09 & Incision and Excision of skull, brain, and cerebral meninges \\
1.2 & Craniotomy and Craniectomy \\
1.24 & Craniotomy \\
1.25 & Craniectomy \\
1.28 & Placement of intacerebral catheters via burr holes \\
1.31 & Incision of Cerebral Meninges \\
8.81 & Linear repair of laceration of eyelid or eyebrow \\
76.72 & Open reduction of malar and zygomatic fracture \\
76.74 & Open reduction of maxillary fracture \\
76.76 & Open reduction of mandibular fracture \\
76.7 & Open reduction of facial fracture \\ \hline
\end{tabular}
\end{table}

Patient-level information contained in both files that was used as linking variables included hospital admission year, length of stay in days, age, gender, race, and indicators for pre-existing chronic conditions such as acute myocardial infarction (AMI), congestive heart failure (CHF), chronic obstructive pulmonary disorder (COPD), diabetes, and hypertension. In addition, the available ICD-9-CM codes on the injury claim were used to classify the type and severity of TBI according to the Barrell Injury Diagnosis Matrix (\cite{Barell2002}), and derive indicators for trauma-related procedures that are commonly associated with TBI.

\newpage

\section{Simulation Results with Day of Birth as Linking Variable}
\begin{singlespace}
\begin{table}[H]
\caption*{Appendix Table 8: Average TPR and PPV across varying error rates for blocking and linking variables when day of birth is available.}
\centering
\begin{tabular}{ccc|ccc|ccl}
 &  &  & \multicolumn{3}{c|}{Average TPR} & \multicolumn{3}{c}{Average PPV} \\
$\epsilon_{Region}$ & $\epsilon_{Income}$ & $\epsilon_{DOB}$&  MLBRL & CIBRL & BRL & MLBRL & CIBRL & BRL \\ \hline
\multirow{9}{*}{0} & \multirow{3}{*}{0} & 0 & 1.00(.002) & 0.99(.009) & 0.99(.004) & 0.87(.020) & 0.88(.025) & 0.76(.014) \\
 &  & 0.2 & 0.85(.010) & 0.85(.011) & 0.82(.009) & 0.80(.022) & 0.80(.027) & 0.63(.010) \\
 &  & 0.4 & 0.70(.016) & 0.70(.018) & 0.63(.015) & 0.69(.022) & 0.69(.022) & 0.48(.013) \\
 & \multirow{3}{*}{0.2} & 0 & 1.00(.002) & 0.81(.041) & 0.99(.004) & 0.88(.024) & 0.88(.024) & 0.76(.013) \\
 &  & 0.2 & 0.84(.011) & 0.70(.039) & 0.83(.008) & 0.79(.036) & 0.78(.025) & 0.63(.009) \\
 &  & 0.4 & 0.69(.017) & 0.56(.036) & 0.63(.017) & 0.68(.031) & 0.65(.029) & 0.48(.016) \\
 & \multirow{3}{*}{0.4} & 0 & 1.00(.002) & 0.76(.055) & 0.99(.003) & 0.88(.024) & 0.88(.030) & 0.75(.010) \\
 &  & 0.2 & 0.85(.011) & 0.65(.044) & 0.82(.010) & 0.81(.029) & 0.79(.023) & 0.62(.010) \\
 &  & 0.4 & 0.69(.017) & 0.53(.041) & 0.63(.013) & 0.69(.024) & 0.66(.025) & 0.48(.010) \\ \hline
\multirow{9}{*}{0.2} & \multirow{3}{*}{0} & 0 & 1.00(.002) & 0.96(.010) & 0.98(.005) & 0.90(.026) & 0.89(.026) & 0.75(.014) \\
 &  & 0.2 & 0.85(.010) & 0.82(.015) & 0.80(.010) & 0.81(.024) & 0.82(.025) & 0.62(.016) \\
 &  & 0.4 & 0.69(.016) & 0.67(.019) & 0.58(.012) & 0.70(.024) & 0.70(.019) & 0.45(.011) \\
 & \multirow{3}{*}{0.2} & 0 & 1.00(.002) & 0.82(.053) & 0.98(.006) & 0.88(.029) & 0.88(.034) & 0.75(.010) \\
 &  & 0.2 & 0.85(.011) & 0.71(.045) & 0.81(.010) & 0.80(.026) & 0.78(.027) & 0.62(.011) \\
 &  & 0.4 & 0.68(.018) & 0.58(.042) & 0.60(.016) & 0.67(.024) & 0.66(.028) & 0.45(.013) \\
 & \multirow{3}{*}{0.4} & 0 & 1.00(.002) & 0.78(.078) & 0.98(.006) & 0.89(.030) & 0.87(.029) & 0.75(.015) \\
 &  & 0.2 & 0.84(.011) & 0.63(.062) & 0.81(.009) & 0.79(.022) & 0.76(.027) & 0.62(.011) \\
 &  & 0.4 & 0.68(.017) & 0.52(.053) & 0.60(.014) & 0.67(.021) & 0.64(.028) & 0.46(.012) \\ \hline
\multirow{9}{*}{0.4} & \multirow{3}{*}{0} & 0 & 1.00(.002) & 0.97(.022) & 0.97(.006) & 0.87(.027) & 0.92(.024) & 0.78(.026) \\
 &  & 0.2 & 0.85(.011) & 0.84(.018) & 0.79(.009) & 0.81(.031) & 0.80(.019) & 0.67(.028) \\
 &  & 0.4 & 0.69(.018) & 0.69(.020) & 0.57(.010) & 0.71(.026) & 0.70(.020) & 0.48(.023) \\
 & \multirow{3}{*}{0.2} & 0 & 1.00(.002) & 0.77(.045) & 0.97(.006) & 0.89(.023) & 0.85(.029) & 0.79(.025) \\
 &  & 0.2 & 0.85(.009) & 0.66(.044) & 0.79(.009) & 0.79(.026) & 0.76(.020) & 0.65(.026) \\
 &  & 0.4 & 0.69(.016) & 0.53(.037) & 0.56(.011) & 0.68(.024) & 0.66(.029) & 0.48(.030) \\
 & \multirow{3}{*}{0.4} & 0 & 1.00(.002) & 0.67(.073) & 0.98(.006) & 0.89(.029) & 0.87(.041) & 0.79(.022) \\
 &  & 0.2 & 0.85(.010) & 0.58(.051) & 0.79(.008) & 0.81(.033) & 0.77(.025) & 0.66(.021) \\
 &  & 0.4 & 0.70(.017) & 0.46(.050) & 0.56(.013) & 0.69(.021) & 0.65(.030) & 0.48(.024) \\ \hline
\end{tabular}
\end{table}
\end{singlespace}

\newpage
\begin{table}[]
\caption*{Appendix Table 9: Average F1 score and Block Accuracy across varying error rates for blocking and linking variables when day of birth is available.}
\centering
\begin{tabular}{ccc|ccc|cc}
 &  &  & \multicolumn{3}{c|}{Average F1 Score} & \multicolumn{2}{c}{Block Accuracy} \\
$\epsilon_{Region}$ & $\epsilon_{Income}$ & $\epsilon_{DOB}$& MLBRL & CIBRL & BRL & MLBRL & CIBRL \\ \hline
\multirow{9}{*}{0} & \multirow{3}{*}{0} & 0 & 0.93(.011) & 0.93(.015) & 0.86(.010) & 1.00(0.00) & 1.00(.010) \\
 &  & 0.2 & 0.82(.013) & 0.82(.016) & 0.71(.009) & 1.00(0.00) & 1.00(.007) \\
 &  & 0.4 & 0.70(.014) & 0.69(.015) & 0.54(.014) & 1.00(0.00) & 1.00(.007) \\
 & \multirow{3}{*}{0.2} & 0 & 0.94(.014) & 0.84(.026) & 0.86(.009) & 1.00(0.00) & 0.81(.041) \\
 &  & 0.2 & 0.82(.021) & 0.73(.025) & 0.72(.008) & 1.00(0.00) & 0.82(.043) \\
 &  & 0.4 & 0.69(.020) & 0.60(.026) & 0.55(.016) & 1.00(0.00) & 0.81(.051) \\
 & \multirow{3}{*}{0.4} & 0 & 0.94(.014) & 0.81(.038) & 0.85(.007) & 1.00(0.00) & 0.76(.055) \\
 &  & 0.2 & 0.83(.015) & 0.71(.030) & 0.71(.010) & 1.00(0.00) & 0.77(.051) \\
 &  & 0.4 & 0.69(.017) & 0.59(.028) & 0.54(.011) & 1.00(0.00) & 0.76(.056) \\ \hline
\multirow{9}{*}{0.2} & \multirow{3}{*}{0} & 0 & 0.95(.014) & 0.92(.014) & 0.85(.010) & 1.00(0.00) & 0.97(.009) \\
 &  & 0.2 & 0.83(.013) & 0.82(.014) & 0.70(.013) & 1.00(0.00) & 0.97(.011) \\
 &  & 0.4 & 0.70(.015) & 0.69(.014) & 0.51(.011) & 1.00(0.00) & 0.97(.014) \\
 & \multirow{3}{*}{0.2} & 0 & 0.93(.016) & 0.85(.035) & 0.85(.008) & 1.00(0.00) & 0.82(.053) \\
 &  & 0.2 & 0.82(.014) & 0.74(.028) & 0.70(.011) & 1.00(0.00) & 0.83(.050) \\
 &  & 0.4 & 0.67(.017) & 0.61(.032) & 0.52(.014) & 1.00(0.00) & 0.84(.050) \\
 & \multirow{3}{*}{0.4} & 0 & 0.94(.017) & 0.82(.048) & 0.85(.010) & 1.00(0.00) & 0.78(.078) \\
 &  & 0.2 & 0.81(.013) & 0.69(.044) & 0.70(.009) & 1.00(0.00) & 0.75(.070) \\
 &  & 0.4 & 0.68(.015) & 0.57(.038) & 0.52(.011) & 1.00(0.00) & 0.76(.072) \\ \hline
\multirow{9}{*}{0.4} & \multirow{3}{*}{0} & 0 & 0.93(.015) & 0.94(.017) & 0.87(.017) & 1.00(0.00) & 0.98(.022) \\
 &  & 0.2 & 0.83(.017) & 0.82(.013) & 0.72(.017) & 1.00(0.00) & 0.98(.017) \\
 &  & 0.4 & 0.70(.017) & 0.70(.016) & 0.52(.014) & 1.00(0.00) & 0.98(.022) \\
 & \multirow{3}{*}{0.2} & 0 & 0.94(.013) & 0.81(.030) & 0.87(.015) & 1.00(0.00) & 0.77(.045) \\
 &  & 0.2 & 0.82(.014) & 0.70(.028) & 0.71(.016) & 1.00(0.00) & 0.77(.048) \\
 &  & 0.4 & 0.69(.017) & 0.59(.026) & 0.51(.019) & 1.00(0.00) & 0.77(.048) \\
 & \multirow{3}{*}{0.4} & 0 & 0.94(.016) & 0.75(.049) & 0.87(.014) & 1.00(0.00) & 0.67(.073) \\
 &  & 0.2 & 0.83(.017) & 0.66(.038) & 0.72(.013) & 1.00(0.00) & 0.67(.060) \\
 &  & 0.4 & 0.69(.015) & 0.54(.039) & 0.52(.018) & 1.00(0.00) & 0.66(.067) \\ \hline
\end{tabular}%
\end{table}

\end{document}